\documentclass[a4paper,fleqn,usenatbib]{mnras}
\pdfoutput=1

\usepackage[T1]{fontenc}
\usepackage{ae,aecompl}

\usepackage{graphicx}	% Including figure files
\usepackage{amsmath}	% Advanced maths commands
\usepackage{amssymb}	% Extra maths symbols
\usepackage{pdflscape}

\title[Circumstellar Disk Lifetimes]{Circumstellar Disk Lifetimes In Numerous Galactic Young Stellar Clusters}

\author[A. J. W. Richert et al.]{
A. J. W. Richert,$^{1}$
K. V. Getman,$^{1}$\thanks{E-mail: kug1@psu.edu (KVG)}
E. D. Feigelson,$^{1}$
M. A. Kuhn,$^{2,3}$\newauthor
P. S. Broos,$^{1}$
M. S. Povich,$^{4}$
M. R. Bate,$^{5}$
G. P. Garmire$^{6}$
\\
$^{1}$Department of Astronomy \& Astrophysics, 525 Davey Laboratory, Pennsylvania State University, University Park PA 16802\\
$^{2}$Instituto de Fisica y Astronomia, Universidad de Valparaiso, Gran Bretana 1111, Playa Ancha, Valparaiso, Chile\\
$^{3}$Millenium Institute of Astrophysics, Av. Vicuna Mackenna 4860, 782-0436 Macul, Santiago, Chile\\
$^{4}$Department of Physics and Astronomy, California State Polytechnic University, 3801 West Temple Ave, Pomona, CA 91768\\
$^{5}$Department of Physics and Astronomy, University of Exeter, Stocker Road, Exeter, Devon EX4 4QL, UK\\
$^{6}$Huntingdon Institute for X-ray Astronomy, LLC, 10677 Franks Road, Huntingdon, PA 16652, USA
}

\date{Accepted for publication in MNRAS, 2018 April 11}

\pubyear{2018}

\begin{document}
\label{firstpage}
\pagerange{\pageref{firstpage}--\pageref{lastpage}}
\maketitle

% Abstract of the paper
\begin{abstract}
Photometric detections of dust circumstellar disks around pre-main sequence (PMS) stars, coupled with estimates of stellar ages, provide constraints on the time available for planet formation. Most previous studies on disk longevity, starting with Haisch, Lada \& Lada (2001), use star samples from PMS clusters but do not consider datasets with homogeneous photometric sensitivities and/or ages placed on a uniform timescale. Here we conduct the largest study to date of the longevity of inner dust disks using X-ray and 1--8~$\micron$  infrared photometry from the MYStIX and SFiNCs projects for 69 young clusters in 32 nearby star-forming regions with ages $t\leq5$~Myr. Cluster ages are derived by combining the empirical Age$_{JX}$ method with PMS evolutionary models, which treat dynamo-generated magnetic fields in different ways. Leveraging X-ray data to identify disk-free objects, we impose similar stellar mass sensitivity limits for disk-bearing and disk-free YSOs while extending the analysis to stellar masses as low as $M\sim0.1$~M$_\odot$.  We find that the disk longevity estimates are strongly affected by the choice of PMS evolutionary model. Assuming a disk fraction of 100\% at zero age, the inferred disk half-life changes significantly, from $t_{1/2}\sim1.3-2$~Myr to $t_{1/2}\sim3.5$~Myr when switching from non-magnetic to magnetic PMS models. In addition, we find no statistically significant evidence that disk fraction varies with stellar mass within the first few Myr of life for stars with masses $<2$~M$_{\odot}$, but our samples may not be complete for more massive stars. The effects of initial disk fraction and star-forming environment are also explored.
\end{abstract}

% Select between one and six entries from the list of approved keywords.
% Don't make up new ones.

%%%%%%%%%%%%%%%%%%%%%%%%%%%%%%%%%%%%%%%%%%%%%%%%%%

%%%%%%%%%%%%%%%%% BODY OF PAPER %%%%%%%%%%%%%%%%%%

\section{Introduction} \label{s:introduction}

The time required to assemble planets in young circumstellar disks remains a key
variable in planet formation theory. Given that planets form out of gas and dust
in young circumstellar disks following protostellar collapse, observed lifetimes
of the gas and dust phases translate into constraints on the time available to
form Jovian and terrestrial planets, respectively \citep{Youdin2005, Lyra2008,
Boss2010}. Measuring disk lifetimes also plays a role in constraining models of
disk evolution more generally, as disk material is depleted by accretion onto
the star, internal and external photoevaporation, disk winds, and planetesimal
and planet formation \citep{Lyndenbell1974, Pringle1981, Bell1997, Armitage2011,
Bai2011, Konigl2011}. Since disk evolution may be dominated by turbulent
viscosity, observationally-derived disk lifetimes are often used to estimate
characteristic viscous $\alpha$ values \citep{Shakura1973}.

Alpha-disk theory allows a disk to be long-lived or short-lived depending on the unknown viscosity \citep{Armitage2011}, therefore in the absence of robust theoretical contraints on effective $\alpha$ values, the actual distribution of longevities for an ensemble of disks must be evaluated observationally. In principle, initial disk masses and disk depletion rates for
individual systems could be used to estimate disk lifetimes, however initial
disk masses cannot be retrospectively determined for individual systems, and
there is no reason to assume that disk dissipation rates are constant
over the long lifetime of a disk. In fact, there is strong evidence that accretion rates are fast during the protostellar phase and slow during later PMS phases, and may be variable on shorter timescales as well \citep{Bouvier1993, Alencar2010, Audard2014, Cody2014}. Therefore, the study of disk longevity typically
relies on disk population statistics rather than observations of individual
systems. 

The earliest empirical study of disk longevity was carried out by
\citet{Strom1989}, who found that the fraction of stars with hot inner accretion disks, detected by
$K$-band excess and H$\alpha$ emission (associated with classical T-Tauri
stars), diminishes significantly for stars older than 3~Myr in the
Taurus--Auriga star-forming region. 

In the seminal study of \citet{Haisch2001b},
the authors plot the fraction of stars showing $L$-band excess, indicating a hot inner dust disk, as a function
of age for several young-to-intermediate age clusters (2.5--30~Myr). They
identify a clear trend wherein disks are depleted over the course of several million years. This
basic methodology of comparing cluster disk fraction with average stellar age
has been adopted by a number of groups for more clusters spanning a wider age
range \citep{Hernandez2008, Mamajek2009, Fedele2010, Bell2013, Ribas2014}. In
some cases, multiple disk indicators are used; \citet{Mamajek2009}, for
instance, uses H$\alpha$ emission, $L$-band excess, 3.6~$\mu$m excess, and
infrared spectral energy distribution (SED) shape as indicators of the presence
of a hot inner disk. There is some evidence that disk longevity depends on host
star spectral type \citep[e.g.,][]{Haisch2001a, Hernandez2005, Carpenter2006, Kennedy2009, Hernandez2010, Luhman2012,
Ribas2015}, with higher-mass stars appearing to shed their disks more quickly.

\citet{Fedele2010} consider the possibility that the dust and gas of a disk do
not perfectly coevolve. They compare the fraction of young stellar objects
(YSOs) showing spectroscopic evidence of accretion onto the star (H$\alpha$
emission) with the fraction showing infrared excess in {\it Spitzer}/IRAC bands
(3.6--8.0~$\mu$m, revealing small grains in the inner several AU of a disk).
Exponential half-lives calculated based on spectroscopic signs of accretion are
slightly shorter than those calculated based on infrared excess. This indicates
that circumstellar gas and dust mostly coevolve, but that disks may retain a
longer-lived dusty component after the gas has been depleted.

One potentially significant limitation of some previous works is the
differing sensitivity limits between disk-bearing and disk-free YSOs.
Disk-bearing YSOs are usually detected through infrared excess, therefore point source
catalogs compiled using infrared photometry are biased toward finding
disk-bearing YSOs. An overestimation of disk fraction will translate into an
overestimation of disk lifetimes. This effect may be particularly important
among lower-mass (i.e., intrinsically fainter) stars, potentially leading to an
apparent mass dependency that is not physical.

Another major impediment to obtaining accurate disk dissipation timescales is the absence of a reliable stellar chronometer. Ages of individual PMS stars as well as age spreads of PMS members in individual clusters and star-forming regions are not accurate due to an interplay of multiple factors, such as photometric variability from accretion and magnetic activity, different accretional histories, binarity, extinction uncertainty, veiling from accretion, scattering and absorption by disks, stellar interiors model uncertainties including inconsistent age predictions for intermediate-mass and lower-mass stars, distance uncertainty, and others \citep[e.g.][]{Preibisch2012, Getman2014, Jeffries2017b}.

Recent empirical evidence points to persistent errors in standard theoretical PMS evolutionary models, both old generation, such as \citet{Baraffe1998, Siess2000}, and new generation, such as \cite{Baraffe2015, Dotter2016, Choi2016}. This emerges independently from: findings of inconsistent ages between intermediate-mass and low-mass stars \citep{Pecaut2016,Fang2017} derived from the Hertzsprung-Russell diagram (HRD) or photometric color-magnitude diagram (CMD); failure of theoretical models to reproduce the observed parameters of stars in eclipsing binaries \citep{Kraus2015}; and disagreement between Li-based and HRD/CMD-based ages \citep{Jeffries2017a}. Specifically, the HRD locations of observed eclipsing binaries suggest that theoretical models underpredict (by $5-20$\%) the stellar radii and overpredict $T_{eff}$ (by $5-10$\%) of low-mass ($M<1$~M$_{\odot}$) PMS and MS stars, a phenomenon referred to as ``radius inflation''. Empirical correlations of inflation with rotation and magnetic activity \citep{Somers2017} suggest that magnetic fields drive radius inflation. Recent attempts to account for magnetic effects include two types of models: global magnetic fields threaded into the stellar interior \citep{Feiden2016} and starspot flux blocking \citep{Somers2015}. Both models lead to changes in the stellar structure that reproduce true radius sizes.

Some of the previous works on cluster disk fraction including \citet{Haisch2001b, Hernandez2008, Mamajek2009, Fedele2010, Ribas2014} employ literature compilations of heterogeneous sets of cluster members and/or heterogeneous estimates of cluster ages. Examples of obvious sources of heterogeneity, which lead to uncertainty and scatter in age estimates, include differing data wavelength ranges, types of data (e.g., photometry versus spectroscopy), age methods (e.g., HRD, CMD, disk fraction, kinematic, etc.),  stellar mass ranges,  ways of transformation between theory and observation, subcluster membership in star-forming regions and others \citep[e.g.,][]{Soderblom2014}. Application of differing PMS evolutionary models to the same set of data would generally lead to systematic shifts in age estimates. None of the aforementioned studies considered the impact of differing PMS models on their disk longevity estimates.

To investigate some of the aforementioned issues in detail, we employ the data from the Massive Young Star-Forming Complex Study in Infrared and X-ray
\citep[MYStIX,][]{Feigelson2013} and Star Formation in Nearby Clouds
\citep[SFiNCs,][]{Getman2017} projects. Both datasets incorporate {\it Chandra} X-ray data to help identify disk-free YSOs in 42 total young
star-forming regions (SFRs). In the current work, we use YSO classifications
(disk-bearing versus disk-free) of X-ray and infrared point sources along with homogeneous sets of cluster ages \citep{Getman2014}
for 69 clusters spread across 32 of the total 42
MYStIX and SFiNCs target regions \citep{Kuhn2014, Getman2018} to study effects of differing PMS evolutionary models, mass, star-forming environments, and initial disk fractions on disk longevity. Ten MYStIX/SFiNCs regions without cluster membership assignments and/or sufficient numbers of stars with available age estimates and disk classes were excluded from our disk fraction analyses (\S \ref{s:membership}).

YSO candidate selection and cluster membership determination are described in
Section~\ref{s:membership}. The classification of disk-bearing and disk-free
YSOs is summarized in Section~\ref{s:yso}. Age estimation for MYStIX and
SFiNCs clusters using multiple PMS evolutionary models is given in Section~\ref{s:ages}. Our strategy for
mitigating the problem of differential mass sensitivities for disk-bearing and
disk-free YSOs is discussed in Section~\ref{s:sensitivity}. The main results for
disk longevity are discussed in Section~\ref{s:results}, including the impacts of different factors (classification of YSOs on disk-bearing and disk-free, assumption of initial disk fraction, choice of PMS model, effects of stellar star-forming environments and mass) on our disk longevity estimates. Further discussion, comparison with previous
literature, and suggestions for future work are presented in
Section~\ref{s:conclusions}.

%%%(In TW Hya, the gas and dust disk structures clearly differ; \citet{Andrews2012}.)

\section{Methods}\label{s:methods}

\subsection{Cluster membership}\label{s:membership}

The MYStIX probable cluster member catalog contains cross-matched X-ray
({\it Chandra}/ACIS), near-infrared (2MASS or UKIDSS), and mid-infrared
({\it Spitzer}/IRAC; 3.6--8.0~$\micron$) point sources. Infrared excess (IRE) selection captures PMS stars with hot inner disks while X-ray selection captures PMS stars with strong magnetic flaring activity. IRE stars are found by comparing the $1-8$~$\micron$ spectral energy distributions to circumstellar disk models \citep{Povich2013} while X-ray stars are found with a naive Bayes classifier that tales a variety of properties into account \citep{Broos2013}. Generally in MYStIX, the {\it Chandra} samples are larger than the IRE samples, but there is often considerable overlap in members identified by the two methods. Detailed discussion of catalog
assembly and membership selection is provided by \citet{Feigelson2013} and other
MYStIX papers \citep{Kuhn2013a, Kuhn2013b, Naylor2013, Povich2013,
Townsley2014}. The full list of $\sim 32000$ MYStIX probable YSO members is given by \citet{Broos2013}; and \citet{Kuhn2014}
identify 142 clusters across 17 MYStIX regions by fitting isothermal
ellipsoids to the star locations. 

The same MYStIX-based X-ray and IR data analysis methods are used for the
reanalysis of the archived {\it Chandra} and {\it Spitzer} data for the nearby 22
SFiNCs SFRs \citep{Getman2017}. Due to the smaller cluster
distances and higher Galactic latitudes of the SFiNCs SFRs compared to
MYStIX ones, the IR counterpart and YSO membership identifications are
achieved using simpler methods than in MYStIX, such as traditional proximity
and decision tree membership classification methods \citep{Getman2017}. The full
list of nearly 8,500 SFiNCs probable YSO members is given by \citet{Getman2017}. 
\citet{Getman2018} identify 52 clusters and 19 unclustered stellar structures
across the 22 SFiNCs SFRs using the methods of
\citet{Kuhn2014}. For our disk longevity analysis, the 19 unclustered stellar
structures are each treated as a single cluster. Throughout the remainder of
this paper, the term ``cluster" will also apply to these 19 unclustered
components.

Sensitivity and completeness levels of the MYStIX and SFiNCs YSO catalogs
vary among the regions due to differing distances, observation exposures, and
absorptions across the fields, as well as due to differing levels of diffuse IR
nebular background. For instance, the very deep X-ray exposure of the nearby ONC
cluster reaches the completeness limits of $\sim 0.1-0.2$~M$_{\odot}$, while the
deep X-ray exposure of the most distant MYStIX region NGC 1893 allows a nearly
complete detection of PMS stars only above $\sim 1-2$~M$_{\odot}$ \citep[see
Figure 1 in][]{Kuhn2015a}. For more distant ($d > 1$~kpc) MYStIX regions, the
2MASS limiting sensitivity of $K_s \sim 14.3$~mag becomes inadequate for
identifying YSO counterparts to {\it Chandra} sources; thus for most of these regions, the 2MASS catalog is complemented by the deeper UKIRT catalog, when available \citep{Feigelson2013}. The {\it Chandra} X-ray-selected and {\it Spitzer} mid-infrared-selected MYStIX/SFiNCs YSO samples generally have different sensitives within individual regions; for instance, an X-ray selected YSO portion is deeper for Be~59 \citep[see Figure 12 in][]{Getman2017}, but a mid-infrared-selected portion is deeper for W~40 \citep[see Figure 8 in][]{Kuhn2015b}.

Due to the omission of {\it Spitzer}-MIPS and far-infrared data, MYStIX and SFiNCs lack the ability to identify some fraction of protostellar objects and transition disk objects (systems with inner disk holes or optically thin inner disks), especially those that were not detected in X-rays. Since the ages of the MYStIX and SFiNCs clusters are estimated based on PMS samples (\S \ref{s:ages}), we are instead interested in characterization of disk fractions for YSO samples, from which the remaining protostars are removed (\S \ref{s:yso}). 

The membership algorithms applied to the MYStIX and SFiNCs X-ray, NIR, and MIR catalogs produce small fractions of false positives. For instance, Table~8 in \citet{Broos2013} shows an excellent (within a few to several percent) agreement between the numbers of the simulated and identified and removed extragalactic and Galactic field contaminants. \citet{Getman2017} estimate that less than a few percent of contaminants, mainly field stars, could be present within the SFiNCs sample of young stellar objects. Getman et al. further compare SFiNCs with previously published YSO catalogs. As an example of the low contamination in SFiNCs membership, here we consider IC~348, one of the richest nearby star-forming regions. Table 9 and Figures 12, 17, and 18 in Getman et al. show that the SFiNCs YSO identification is in good agreement with the recent optical/infrared spectroscopic/photometric YSO catalog of \citep[][hereafter Lu16]{Luhman2016}. Out of the 478 Lu16 YSOs, 77\% are identified by SFiNCs. Half of the remaining 23\% (generally IR brighter) lie outside the SFiNCs X-ray fields, and the other half (IR weaker) are very low-mass stellar and brown dwarf candidates ($M \la 0.1$~M$_{\odot}$) undetectable in the SFiNCs X-ray exposures. On the other hand, additional to Lu16, SFiNCs identifies 29 new YSOs; of those half are disk-free and half are disk-bearing; over two-thirds are X-ray detected. The fact that the vast majority of these are not distributed randomly across the SFiNCs X-ray field but are rather spatially concentrated in the southern part of the field, right outside the primary membership area of Lu16 ($r=14\arcmin$; their Figure 1), and have IR/X-ray colors consistent with those of the other YSOs, gives confidence that these are real YSOs and not source contaminants.

Only a portion of the full MYStIX and SFiNCs samples can be used for our effort to understand disk evolution. Specifically, we require clusters with available $Age_{JX}$ estimates (Section \ref{s:ages}) and at least 10 disk-bearing and 10 disk-free YSOs within each cluster \textit{prior} to the imposition of mass cuts described in Section~\ref{s:sensitivity}. The resulting subsample consists of 7,100 MYStIX YSOs in 34 clusters, and 5,834 SFiNCs YSOs in 35 clusters. These numbers of YSOs are further reduced after the imposition of mass cuts (Section~\ref{s:sensitivity}); and the numbers vary among the four different membership permutations considered in the current study (Section~\ref{s:yso}). The final YSO numbers are listed in Tables \ref{t:clusters} and \ref{t:longevity}.

\begin{figure}
\includegraphics[angle=0.,width=75mm]{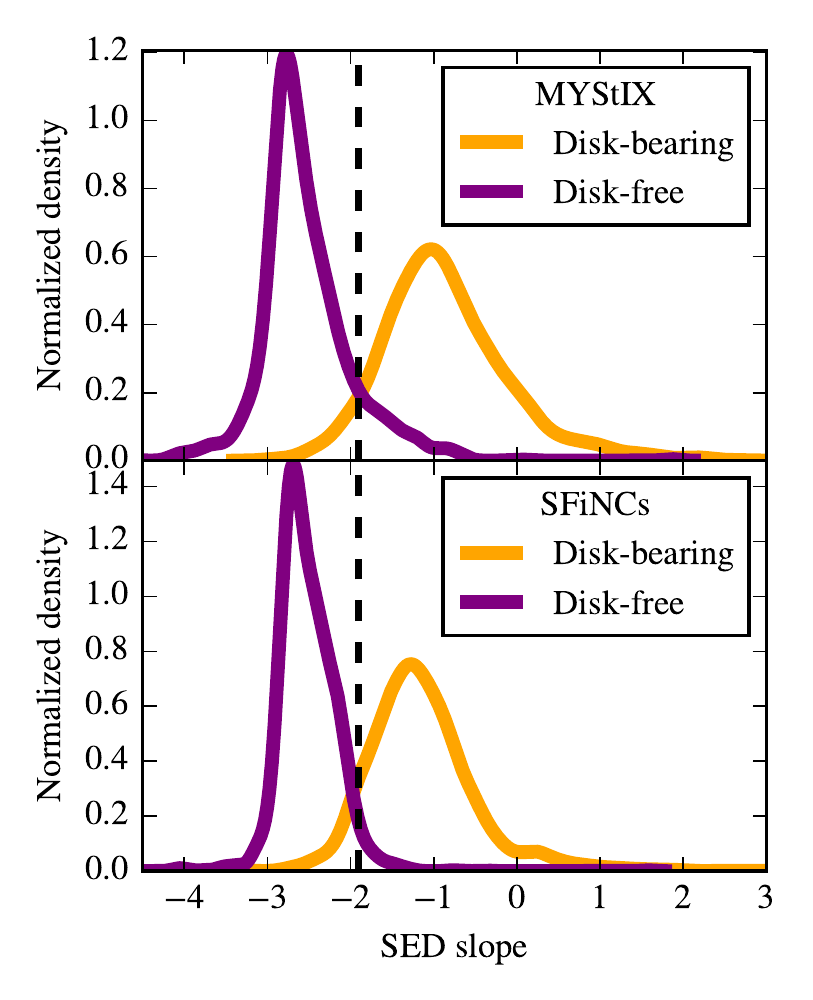}
\caption{Kernel density estimation of SED slopes $\alpha_{IRAC}$ for MYStIX
(upper panel) and SFiNCs (lower panel) YSOs.}
\label{f:pretty_slope_KDE}
\end{figure}

\subsection{YSO classification}\label{s:yso}

There is no consensus on criteria for discriminating between disk-bearing and disk-free stars. We apply and compare two schemes for doing so. In the first, we use the
classifications found in the MYStIX and SFiNCs catalogs derived in the
following ways. For MYStIX, \citet{Povich2013} classify YSOs by fitting
$JHK_s$ and {\it Spitzer}/IRAC photometry with the model spectral energy
distributions of \citet[][disks]{Robitaille2006} and \citet[][stellar
photospheres]{Castelli2004}, and removing numerous contaminating sources
(extragalactic objects, asymptotic giant branch stars, nebular knots, and
unrelated YSOs) through additional infrared color cuts and spatial clustering
analyses. X-ray detections, an indicator of youth, are required for disk-free
YSOs in order to exclude field stars. SFiNCs regions are out of the Galactic
plane and therefore suffer from less contamination than MYStIX regions.
\citet{Getman2017} therefore classify SFiNCs YSOs using simpler procedures,
namely, the {\it Spitzer}/IRAC color--color diagram scheme of \citet{Gutermuth2009}
combined with the SED-based analysis of \citet{Getman2012}. We refer to this
first set of YSO classifications collectively as ``catalog class."

Our second YSO classification scheme is based on
apparent (non-dereddened) {\it Spitzer}/IRAC SED slopes, $\alpha_{IRAC} = d \log(\lambda F_{\lambda})/d \log(\lambda)$, measured in the IRAC wavelength range from 3.6 to 8.0~$\mu$m. Available magnitudes in the 3.6, 4.5, 5.8, and 8.0~$\mu$m bands were used for the calculation of 100\%, 100\%, 64\%, and 49\% of the SED slopes, respectively. We use a critical value of $\alpha_{IRAC}=-1.9$ to distinguish between
disk-bearing and disk-free YSOs. In Fig.~\ref{f:pretty_slope_KDE}, we plot
kernel density distributions of $\alpha_{IRAC}$ for MYStIX and SFiNCs,
separately for disk-bearing and disk-free objects as determined by catalog class
(discussed in the previous paragraph). In both figures, the crossover point
between disk-bearing and disk-free is close to $-1.9$. For MYStIX, 91\% of
disk-bearing YSOs have $\alpha_{IRAC}>-1.9$, while 86\% of disk-free YSOs have
$\alpha_{IRAC}<-1.9$. For SFiNCs, 91\% of disk-bearing YSOs have
$\alpha_{IRAC}>-1.9$, while 96\% of SFiNCs disk-free YSOs have
$\alpha_{IRAC}<-1.9$. The separation between disk-bearing and disk-free objects
therefore appears to be consistent between MYStIX and SFiNCs. The two YSO classification schemes disagree for $<14$\% ($<9$\%) of the MYStIX (SFiNCs) stars.

For both YSO classification schemes, we repeat our analysis (presented in
\S~\ref{s:results}) while excluding probable protostars. We define probable
protostars as being those objects with (non-dereddened) $\alpha_{IRAC}>0$.

We utilize both of the aforementioned schemes not only to compare these
approaches---model SED-fitting, color--color diagrams, and infrared
slope-based classifications are all widely used in the literature---but also due
to the fact that model SED fitting and color--color based schemes yield
many ambiguous classifications, including due to transition disks. The latter typically show weak or no IR excesses in the IRAC bands, adding to the uncertainties in disk classification and resulting disk fraction.

\subsection{Cluster ages} \label{s:ages}

MYStIX and SFiNCs cluster ages are determined using the $Age_{JX}$ method
described by \citet{Getman2014}. $Age_{JX}$ is applicable only to low-mass PMS stars ($M<1.2$~M$_{\odot}$ assuming the \citet{Siess2000} age scale) with reliable measurements of the intrinsic X-ray luminosity and near-infrared $JHK_s$ photometry. X-ray luminosities ($L_X$) specify stellar mass
according to the empirical PMS correlation seen in the Taurus region
\citep{Telleschi2007}. $J$-band luminosities and mass estimates track with PMS
evolutionary models, providing stellar ages. This yields homogeneous median age
estimates for all 69 clusters used in the
current analysis. To investigate how the choice of theoretical PMS evolutionary models affects disk dissipation timescales, the $Age_{JX}$ method is applied to the MYStIX and SFiNCs YSOs using a number of different models.

We start with four different sets of stellar evolutionary models: \citet{Siess2000}[hereafter Siess00]; \citet{Baraffe2015}[hereafter Baraffe15]; \citet{Dotter2016} and \citet{Choi2016}[hereafter MIST]; and \citet{Feiden2015} and \citet{Feiden2016}[hereafter Feiden16]. A quick examination of these models' evolutionary isochrones placed on the $L_{bol}$ - $T_{eff}$ diagram suggests that, within the locus of the MYStIX/SFiNCs YSOs, the predictions of the Baraffe15 and MIST models are in good agreement with each other (figure not shown). Due to a poorer-sampled published model grid, compared to MIST, the Baraffe15 model is omitted from further consideration. Compared to Siess00, newer generations of standard evolutionary models, such as Baraffe15 and MIST, with improved microphysics (including updated solar abundance scale, linelists, atmospheric convection parameters) predict systematically younger ages.

The $Age_{JX}$ estimates for MYStIX and SFiNCs clusters based on the Siess00 model are already reported in \citet{Getman2014, Getman2018}. Here we refer to these ages as $Age_{JX}$-Siess00. Next, we recalculate cluster ages using MIST as an underlying evolutionary model and following the same $Age_{JX}$ procedure detailed in \citet{Getman2014}. Here we refer to these ages as $Age_{JX}$-MIST. Briefly, the $L_X$-Mass relationship of \citet{Telleschi2007} is recalibrated by comparing the ($T_{eff},L_{bol}$) measurements for Taurus X-ray emitting PMS stars \citep{Gudel2007} to the MIST models. The X-ray luminosities of all $Age_{JX}$ stars in MYStIX and SFiNCs are then converted to stellar masses. The differences between the resulting MIST and Siess00 based stellar masses are no more than 15\%. Comparison of absolute $J$-band magnitudes and masses with the MIST evolutionary tracks yields ages for individual stars. Cluster ages are then calculated as median ages of all $Age_{JX}$ stellar cluster members. Statistical errors on cluster ages are calculated as 68\% confidence intervals using nonparametric bootstrap resampling. The bootstrap case resampling takes into account any forms of observed scatter; thus all sources of the scatter including the uncertainties on individual source extinctions, stellar masses, and local distances within the cluster \citep[described in \S\S 3.3 and 5 in][]{Getman2014} are treated naturally. Bootstrap does not treat ``systematic'' uncertainties, such as the uncertainty in the knowledge of PMS evolutionary models or uncertainty on the distance from the Sun to the region \citep[the latter is discussed in \S 5 in][]{Getman2014}; such effects globally shift age scales.

\begin{figure}
\includegraphics[width=75mm]{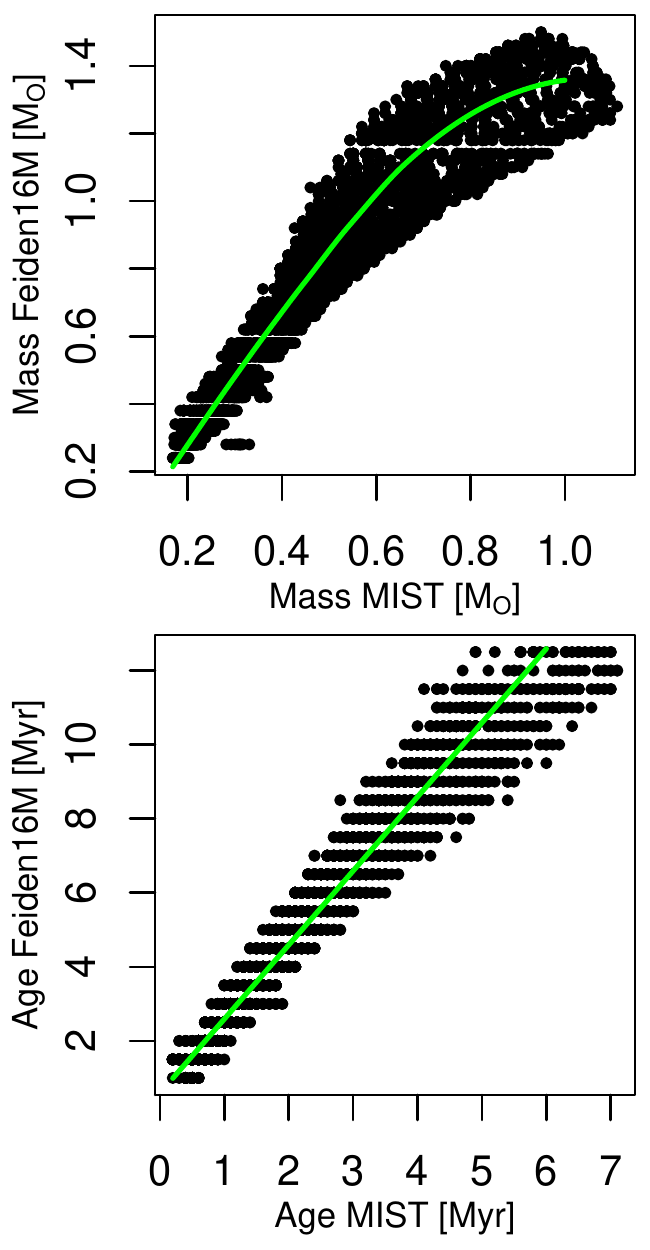}
\caption{Comparison of the non-magnetic MIST \citep{Dotter2016, Choi2016} and magnetic Feiden16 \citep{Feiden2015, Feiden2016} PMS evolutionary models. Stellar masses (upper panel) and stellar ages (lower panel) resulted from the model fitting of simulated stars on the $(L_{bol},T_{eff})$ diagram. The green curves are polynomial and linear (standard major axis) regression fits for the mass and age data, respectively. The polynomial and linear fits were obtained using the {\it R}\citep{R2014} functions {\it loess} and {\it lmodel2}, respectively.}
\label{f:feiden16m_vs_mist}
\end{figure}

The Dartmouth stellar evolution model grid \citep{Dotter2008} extended to the PMS phase is used as a basis for the Feiden16 models \citep{Feiden2015}. Two versions of the Feiden16 models are provided, ``non-magnetic'' and ``magnetic'' \citep{Feiden2016}. The published model grid is rather poorly sampled ($\Delta t = 0.5$~Myr) with no predictions for photometric magnitudes. Thus, the usage of the $Age_{JX}$ method is not feasible here; and instead we opt for a simple age rescaling technique. In this approach, 5,000 young stars within the parameter ranges of the MYStIX/SFiNCs-$Age_{JX}$ stars are simulated on the $L_{bol}$ versus $T_{eff}$ diagram. Their ages are then derived using both the MIST and Feiden16 models. The resulting Feiden16 versus MIST age relationships are approximated by linear functions using standard major axis regression: $t_{F16,non-magnetic} = 0.35 + 0.94 \times t_{MIST}$ and $t_{F16,magnetic} = 0.59 + 2.00 \times t_{MIST}$. 

Fig. \ref{f:feiden16m_vs_mist} exemplifies comparison of the standard MIST and ``magnetic'' Feiden16 models. Compared to MIST, the Feiden16 ``magnetic'' models predict systematically higher (by a factor of 1.6) masses and older (by a factor of $2-3$) ages. Following the above equations, the $Age_{JX}$-MIST ages for the MYStIX/SFiNCs clusters are then transformed to $Age$-Feiden16 cluster ages. The ``non-magnetic'' Feiden16 cluster ages are omitted from further consideration because they only slightly differ from the MIST ages ($<30$\% for 1~Myr and $< 10$\% for $>2$~Myr clusters) and are systematically lower than the Siess00 ages. In contrast, the ``magnetic'' Feiden16 cluster ages appear significantly higher than the MIST and Siess00 ages; we refer to these hereafter as $Age$-Feiden16M.

For all MYStIX/SFiNCs clusters, three types of cluster ages are reported in Table~\ref{t:clusters}: old generation standard Siess00 ($Age_{JX}$-Siess00), new generation standard MIST ($Age_{JX}$-MIST), and new generation magnetic Feiden16 ($Age$-Feiden16M).

\subsection{Stellar mass sensitivity}\label{s:sensitivity}

Given that disks around YSOs are detected using infrared excess, for a given
infrared limiting magnitude, disk-bearing YSOs will be more readily detected
than disk-free ones. Ignoring this sensitivity difference could lead to an
erroneously high estimate of disk fraction---and therefore disk longevity---that
reflects the underdetection of low-mass, disk-free YSOs. Note that this effect
will manifest even if disk longevity does not vary with host star mass, and
could wrongly lead to an {\it apparent} dependency of disk fraction on
stellar mass.

In order to ensure that disk fractions are calculated within a similar range of
stellar mass for each cluster, we choose a sensitivity limit for each cluster
in the following way. Approximate stellar mass estimates are obtained based on
star locations in the {$J$} versus $J-H$ color--magnitude diagram
and theoretical stellar model tracks derived by \citet[][for
$M_{\star} <7$~M$_{\odot}$]{Siess2000} and \citet[][for $M_{\star} >7$~M$_{\odot}$
stars]{Bressan2012}. The reddening law of \citet{Rieke1985} is used to deredden
the star positions to the intrinsic pre-main sequence model colors, assuming a
single age ($Age_{JX}$-Siess00) for all the stellar members of a cluster. These mass
estimates are subject to significant uncertainties and may be incompatible with
individual masses obtained by other methods such as optical spectroscopy
\citep{Kuhn2010}. However, the analysis of disk longevity as a function of
stellar mass presented in Section~\ref{s:massbins} makes use of large stellar
mass bins, so precise estimates of mass for individual objects are not
required.

\begin{figure}
\includegraphics[width=75mm]{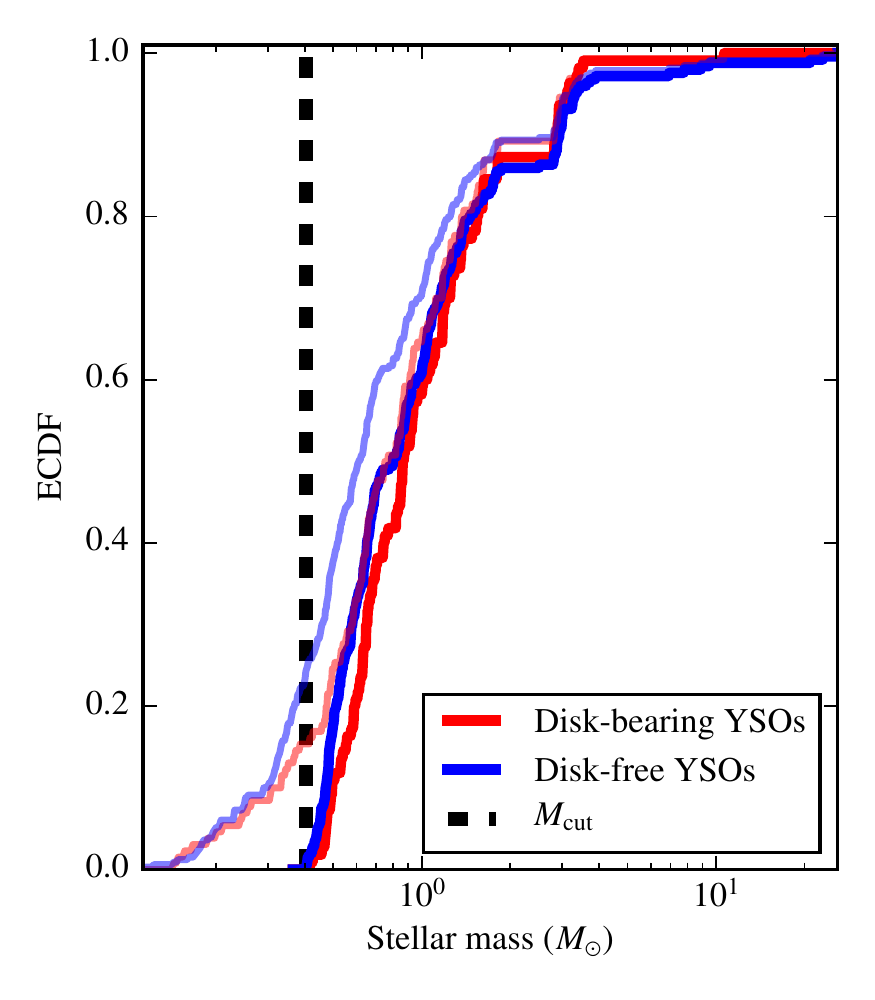}
\caption{Comparison of stellar mass cumulative distributions for disk-bearing
and disk-free YSOs in Rosette cluster E (NGC~2244), based on catalog YSO
classifications. The thinner (translucent) and thicker (solid) lines indicate
the mass distributions before and after the imposition of $M_{cut}$,
respectively.}
\label{f:examplemassECDF}
\end{figure}

Once stellar masses have been calculated, we compare the stellar mass
distributions of disk-bearing and disk-free YSOs in each of our 69
clusters. For each cluster, we iteratively remove the lowest-mass YSO in the
combined disk-bearing and disk-free sample until the Kolmogorov--Smirnov
two-sample $p$-value exceeds 0.1. This yields a minimum mass cutoff $M_{cut}$
for each cluster in order to ensure similar mass distributions and mass
sensitivities for disk-bearing and disk-free YSOs. As mentioned in the next
subsection, for many clusters, no YSOs are removed because they already show
similar mass completeness for their disk-bearing and disk-free YSOs.
Fig.~\ref{f:examplemassECDF} shows empirical cumulative distribution functions
(ECDFs) of stellar mass for disk-bearing and disk-free objects in Rosette Nebula
cluster E \citep[NGC~2244;][]{Kuhn2014}, illustrating the derivation of $M_{cut}$ and the effect
of imposing it. For MYStIX clusters, typical values of $M_{cut}$ typically range from 0.1 to 0.2~$M_\odot$, while for SFiNCs clusters, minimum stellar masses in the sample reach $\sim0.1~M_\odot$, but no $M_{cut}$ was applied due to the similar mass distributions of disk-bearing and disk-free stars.

Resulting subsamples of disk-bearing and disk-free stars, with similarly shaped mass ECDFs, are not affected by the choice of PMS evolutionary models insofar as both disk-bearing and disk-free samples are subject to the same mass transformation when transitioning among different PMS models. The mass scale itself changes; it is roughly similar between the Siess00 and MIST models, but shifts to higher values upon switching to the Feiden16M model (\S \ref{s:ages}).

\subsection{Summary of cluster data}\label{s:clustersummary}

Table~\ref{t:clusters} presents the cluster data.

\newpage 

\begin{table}\small
\centering
 \begin{minipage}{180mm}
 \caption{Properties of 69 MYStIX and SFiNCs clusters. This version of the table lists 69 entries (one per cluster) corresponding to the membership case with $\alpha_{IRAC}$-based classification and probable protostars included. This table is available in its entirety ($69 \times 4$ entries) in the Supplementary Materials. That is, the on-line table version gives four entries per cluster, one for each of the four membership permutations yielded by using two different YSO classification schemes (YSO Classes = Catalog and YSO Classes = $\alpha_{IRAC}$) and including and excluding probable protostars. Here, Column 1: Cluster of interest. Column 2: Distances from the Sun, taken from \citet[MYStIX;][]{Feigelson2013} and \citet[SFiNCs;][]{Getman2017}. Columns 3-5: Cluster ages using three different PMS evolutionary models. Columns 6-7: Minimum mass cut-off and median stellar mass in a cluster (based on the Siess00 age scale). Columns 8-9: Numbers of disk-bearing ($N_{disk}$) and disk-free ($N_{nodisk}$) YSOs after the imposition of $M_{cut}$. Column 10: Inferred disk fraction $f_{disk} = N_{disk}/(N_{disk}+N_{nodisk})$.}
 \label{t:clusters}
%% \begin{tabular}{@{}ccccccccccccccc}
 \begin{tabular}{@{\vline }c@{ \vline }c@{ \vline }c@{ \vline }c@{ \vline }c@{ \vline }c@{ \vline }c@{ \vline }c@{ \vline }c@{ \vline }c@{ \vline }}
\cline{1-10}
&&&&&&&&\\ 
%\hline
Region/Cluster & D & $Age_{JX}$ & $Age_{JX}$ &
$Age$ & $M_{cut}$ & Median mass &
$N_{disk}$ & $N_{nodisk}$ & $f_{disk}$\\
& &  Siess00      &  MIST          & Feiden16M &  Siess00 & Siess00     &        &          &                 \\
     & (kpc)          & (Myr)         &  (Myr)         & (Myr)     &  (M$_{\odot}$) &   (M$_{\odot}$)   &        &          &                 \\
   (1)         & (2)         &  (3)         & (4)     &  (5) &   (6)   &   (7)      &   (8)       &    (9)  & (10)            \\
\cline{1-10}
&&&&&&&&\\
\hline
\multicolumn{10}{c}{MYStIX} \\
\hline
Eagle/A  & 1.75  & 2.4$\pm$1.0  & 2.0$\pm$0.8 & 4.5 &    0.95  &  1.24  &  3  &  13  &  $0.19^{+0.12}_{-0.19}$ \\
Eagle/B  & 1.75  & 2.1$\pm$0.1  & 1.5$\pm$0.1 & 3.6 &    0.25  &  1.17  &  295  &  348  &  $0.46^{+0.04}_{-0.04}$ \\
Eagle/D  & 1.75  & 2.5$\pm$0.2  & 1.6$\pm$0.2 & 3.8 &    0.16  &  1.47  &  151  &  153  &  $0.50^{+0.06}_{-0.06}$ \\
Flame/A  & 0.414 & 0.8$\pm$0.2  & 0.4$\pm$0.1 & 1.5    &    0.10:  &  0.38  &  101  &  41  &  $0.71^{+0.08}_{-0.07}$ \\
Lagoon/A  & 1.3  & 2.2$\pm$0.2  & 1.7$\pm$0.1 & 4.0   &    0.18:  &  0.69  &  16  &  11  &  $0.59^{+0.19}_{-0.16}$ \\
Lagoon/C  & 1.3  & 1.6$\pm$0.2  & 1.0$\pm$0.3 & 2.7   &    0.11:  &  1.13  &  19  &  12  &  $0.61^{+0.17}_{-0.15}$ \\
Lagoon/E  & 1.3  & 1.9$\pm$0.2  & 1.4$\pm$0.2 & 3.4   &    0.11:  &  1.06  &  34  &  44  &  $0.44^{+0.10}_{-0.11}$ \\
Lagoon/F  & 1.3  & 2.3$\pm$0.1  & 1.8$\pm$0.2 & 4.2   &    0.12  &  0.96  &  108  &  153  &  $0.41^{+0.06}_{-0.06}$ \\
Lagoon/H  & 1.3  & 2.1$\pm$0.4  & 1.3$\pm$0.3 & 3.2   &    0.10:  &  1.25  &  47  &  43  &  $0.52^{+0.10}_{-0.10}$ \\
Lagoon/I  & 1.3  & 2.1$\pm$0.2  & 1.5$\pm$0.2 & 3.6   &    0.10:  &  0.97  &  79  &  85  &  $0.48^{+0.08}_{-0.08}$ \\
Lagoon/J  & 1.3  & 2.7$\pm$0.2  & 1.9$\pm$0.3 & 4.4   &    0.11:  &  1.28  &  19  &  32  &  $0.37^{+0.12}_{-0.14}$ \\
Lagoon/K  & 1.3  & 1.4$\pm$0.2  & 1.0$\pm$0.2 & 2.6   &    0.11:  &  1.19  &  60  &  28  &  $0.68^{+0.10}_{-0.09}$ \\
M~17/D  &  2.0   & 1.1$\pm$0.2  & 0.7$\pm$0.1 & 2.0     &    0.12:  &  3.68  &  21  &  14  &  $0.60^{+0.16}_{-0.14}$ \\
NGC~1893/A  & 3.6 & 3.5$\pm$1.0  & 2.7$\pm$0.9 & 5.9 &    0.19:  &  1.81  &  21  &  39  &  $0.35^{+0.11}_{-0.13}$ \\
NGC~1893/B  & 3.6 & 2.6$\pm$0.3  & 2.0$\pm$0.3 & 4.7 &    0.09:  &  1.15  &  56  &  68  &  $0.45^{+0.08}_{-0.09}$ \\
NGC~1893/I  & 3.6 & 2.8$\pm$0.6  & 1.9$\pm$0.3 & 4.3 &    0.12:  &  1.36  &  66  &  51  &  $0.56^{+0.09}_{-0.09}$ \\
NGC~2264/E  & 0.913 & 3.2$\pm$0.5  & 2.4$\pm$0.7 & 5.4 &    0.10:  &  0.58  &  17  &  58  &  $0.23^{+0.08}_{-0.11}$ \\
NGC~2264/J  & 0.913 & 1.6$\pm$0.7  & 1.2$\pm$0.6 & 3.0 &    0.10:  &  0.75  &  35  &  15  &  $0.70^{+0.14}_{-0.11}$ \\
NGC~2264/K  & 0.913 & 2.2$\pm$0.3  & 1.5$\pm$0.2 & 3.6 &    0.09:  &  0.66  &  39  &  31  &  $0.56^{+0.12}_{-0.11}$ \\
NGC~2362/B  & 1.48 &  2.9$\pm$0.2  & 2.1$\pm$0.2 & 4.8 &    0.09:  &  0.50  &  23  &  177  &  $0.12^{+0.04}_{-0.05}$ \\
NGC~6334/B  & 1.7  & 2.3$\pm$0.4  & 1.8$\pm$0.5 & 4.1 &    0.17:  &  1.92  &  26  &  14  &  $0.65^{+0.15}_{-0.13}$ \\
NGC~6334/J  & 1.7  & 1.5$\pm$0.4  & 0.9$\pm$0.4 & 2.3 &    0.11:  &  3.85  &  20  &  4  &  $0.83^{+0.19}_{-0.10}$ \\
NGC~6334/L  & 1.7  & 0.7$\pm$0.3  & 0.4$\pm$0.2 & 1.4 &    0.39  &  1.60  &  20  &  1  &  $0.95^{+0.18}_{-0.04}$ \\
NGC~6357/A  & 1.7  & 1.4$\pm$0.1  & 0.9$\pm$0.1 & 2.4 &    0.13:  &  1.58  &  60  &  64  &  $0.48^{+0.09}_{-0.09}$ \\
NGC~6357/B  & 1.7  & 1.4$\pm$0.2  & 1.0$\pm$0.2 & 2.6 &    0.09:  &  1.46  &  94  &  62  &  $0.60^{+0.08}_{-0.07}$ \\
NGC~6357/C  & 1.7  & 1.2$\pm$0.3  & 0.9$\pm$0.2 & 2.4 &    0.41  &  1.38  &  71  &  67  &  $0.51^{+0.08}_{-0.08}$ \\
NGC~6357/E  & 1.7  & 1.4$\pm$0.4  & 0.9$\pm$0.3 & 2.3 &    0.14:  &  1.86  &  38  &  26  &  $0.59^{+0.12}_{-0.11}$ \\
NGC~6357/F  & 1.7  & 1.5$\pm$0.2  & 1.0$\pm$0.2 & 2.6 &    0.11:  &  1.79  &  90  &  72  &  $0.56^{+0.08}_{-0.07}$ \\
RCW~36/A  &  0.7   &0.9$\pm$0.1  & 0.5$\pm$0.1 & 1.6 &    0.09:  &  0.35  &  105  &  24  &  $0.81^{+0.08}_{-0.06}$ \\
Rosette/E  &  1.33 &3.0$\pm$0.2  & 2.3$\pm$0.1 & 5.3 &    0.21  &  0.66  &  130  &  335  &  $0.28^{+0.04}_{-0.04}$ \\
Rosette/L  &  1.33 &2.7$\pm$0.7  & 1.9$\pm$1.0 & 4.3 &    0.10:  &  0.81  &  136  &  95  &  $0.59^{+0.06}_{-0.06}$ \\
Rosette/M  &  1.33 &1.9$\pm$0.4  & 1.4$\pm$0.3 & 3.3 &    0.11:  &  0.81  &  43  &  8  &  $0.84^{+0.12}_{-0.08}$ \\
Rosette/N  &  1.33 &1.3$\pm$1.4  & 0.7$\pm$1.0 & 2.0 &    0.12:  &  0.68  &  14  &  15  &  $0.48^{+0.17}_{-0.17}$ \\
W~40/A  &  0.5     &0.8$\pm$0.1  & 0.4$\pm$0.1 & 1.5 &    0.10:  &  0.53  &  123  &  33  &  $0.79^{+0.07}_{-0.06}$ \\
&&&&&&&&\\
\cline{1-10} 
\end{tabular}
\end{minipage}
\end{table}
\clearpage
\newpage

\begin{table}\small
\centering
 \begin{minipage}{180mm}
%% \begin{tabular}{@{}ccccccccccccccc}
 \begin{tabular}{@{\vline }c@{ \vline }c@{ \vline }c@{ \vline }c@{ \vline }c@{ \vline }c@{ \vline }c@{ \vline }c@{ \vline }c@{ \vline }c@{ \vline }}
\cline{1-10}
&&&&&&&&\\ 
%\hline
Region/Cluster & D & $Age_{JX}$ & $Age_{JX}$ &
$Age$ & $M_{cut}$ & Median mass &
$N_{disk}$ & $N_{nodisk}$ & $f_{disk}$\\
& & Siess00      &  MIST          & Feiden16M &  Siess00 & Siess00     &        &          &                 \\
     & (kpc)          & (Myr)         &  (Myr)         & (Myr)     &  (M$_{\odot}$) &   (M$_{\odot}$)   &        &          &                 \\
   (1)         & (2)         &  (3)         & (4)     &  (5) &   (6)   &   (7)      &   (8)       &    (9)  & (10)            \\
\cline{1-10}
&&&&&&&&\\
\hline
\multicolumn{10}{c}{SFiNCs} \\
\hline
Be~59/A  & 0.9 &  1.8$\pm$0.2  & 1.4$\pm$0.2 & 3.3 &    0.10:  &  0.78  &  149  &  152  &  $0.50^{+0.06}_{-0.06}$ \\
Be~59/B  & 0.9 & 2.2$\pm$0.4  & 1.6$\pm$0.3 & 3.8 &    0.75  &  1.59  &  35  &  60  &  $0.37^{+0.09}_{-0.10}$ \\
Cep~A/A  & 0.7 & 1.4$\pm$0.3  & 1.0$\pm$0.2 & 2.6 &    0.10:  &  0.38  &  50  &  27  &  $0.65^{+0.11}_{-0.10}$ \\
Cep~A/U  & 0.7 & 2.0$\pm$0.3  & 1.4$\pm$0.3 & 3.3 &    0.10:  &  0.43  &  29  &  57  &  $0.34^{+0.09}_{-0.10}$ \\
Cep~C/U  & 0.7 & 2.2$\pm$0.9  & 2.0$\pm$0.6 & 4.6 &    0.10:  &  0.47  &  26  &  33  &  $0.44^{+0.12}_{-0.13}$ \\
Cep~OB3b/A & 0.7 &  2.2$\pm$0.2  & 1.9$\pm$0.2 & 4.4 &    0.09:  &  0.36  &  201  &  195  &  $0.51^{+0.05}_{-0.05}$ \\
Cep~OB3b/C & 0.7 & 2.4$\pm$0.1  & 2.0$\pm$0.1 & 4.6 &    0.09:  &  0.44  &  284  &  324  &  $0.47^{+0.04}_{-0.04}$ \\
Cep~OB3b/U & 0.7 & 3.4$\pm$0.4  & 2.5$\pm$0.6 & 5.6 &    0.10:  &  0.42  &  33  &  48  &  $0.41^{+0.10}_{-0.11}$ \\
GGD~12-15/A & 0.83 &  0.6$\pm$0.6  & 0.4$\pm$0.6 & 1.4 &    0.10:  &  0.34  &  44  &  11  &  $0.80^{+0.12}_{-0.08}$ \\
GGD~12-15/U & 0.83 & 2.5$\pm$0.5  & 2.2$\pm$0.4 & 5.0 &    0.11:  &  0.43  &  31  &  57  &  $0.35^{+0.09}_{-0.10}$ \\
IC~348/B  &  0.3  & 2.5$\pm$0.1  & 2.0$\pm$0.2 & 4.6 &    0.09:  &  0.36  &  91  &  129  &  $0.41^{+0.06}_{-0.07}$ \\
IC~348/U  &  0.3  & 3.8$\pm$0.4  & 3.4$\pm$0.6 & 7.4 &    0.11:  &  0.35  &  11  &  27  &  $0.29^{+0.12}_{-0.16}$ \\
IC~5146/B  & 0.8  & 1.5$\pm$0.2  & 1.2$\pm$0.2 & 3.1 &    0.10:  &  0.57  &  88  &  37  &  $0.70^{+0.09}_{-0.07}$ \\
IC~5146/U  & 0.8  & 2.6$\pm$0.5  & 2.0$\pm$0.3 & 4.6 &    0.30  &  0.67  &  37  &  34  &  $0.52^{+0.11}_{-0.11}$ \\
IRAS~00013+681/A  & 0.9 &  1.8$\pm$1.8  & 1.2$\pm$1.4 & 3.0 &    0.18  &  0.68  &  24  &  9  &  $0.73^{+0.17}_{-0.12}$ \\
IRAS~20050+2720/U  & 0.7 & 3.3$\pm$0.4  & 2.6$\pm$0.4 & 5.8 &    0.57  &  0.99  &  8  &  20  &  $0.29^{+0.13}_{-0.18}$ \\
LkH$\alpha$~101/A  & 0.51 & 1.5$\pm$0.3  & 1.0$\pm$0.3 & 2.6 &    0.09:  &  0.37  &  78  &  62  &  $0.56^{+0.08}_{-0.08}$ \\
LkH$\alpha$~101/U  & 0.51 & 2.2$\pm$0.6  & 1.7$\pm$0.5 & 4.0 &    0.10:  &  0.34  &  23  &  22  &  $0.51^{+0.14}_{-0.14}$ \\
Mon~R2/A  & 0.83 & 1.2$\pm$0.1  & 0.9$\pm$0.1 & 2.4 &    0.09:  &  0.28  &  65  &  13  &  $0.83^{+0.10}_{-0.07}$ \\
Mon~R2/U  & 0.83 & 1.7$\pm$0.2  & 1.4$\pm$0.2 & 3.3 &    0.09  &  0.47  &  134  &  74  &  $0.64^{+0.07}_{-0.06}$ \\
NGC~1333/A  & 0.235 & 2.5$\pm$1.2  & 1.2$\pm$1.5 & 3.0 &    0.09:  &  0.17  &  16  &  6  &  $0.73^{+0.21}_{-0.14}$ \\
NGC~1333/B  & 0.235 & 1.7$\pm$0.3  & 1.3$\pm$0.3 & 3.2 &    0.09:  &  0.18  &  29  &  15  &  $0.66^{+0.15}_{-0.12}$ \\
NGC~2068/B  & 0.414 & 1.2$\pm$0.2  & 0.7$\pm$0.1 & 2.0 &    0.10:  &  0.43  &  57  &  34  &  $0.63^{+0.10}_{-0.09}$ \\
NGC~2068/D  & 0.414 & 1.0$\pm$0.3  & 0.6$\pm$0.2 & 1.8 &    0.11:  &  0.59  &  23  &  11  &  $0.68^{+0.17}_{-0.13}$ \\
NGC~2068/U  & 0.414 & 2.3$\pm$0.4  & 1.4$\pm$0.5 & 3.4 &    0.09:  &  0.41  &  37  &  40  &  $0.48^{+0.11}_{-0.11}$ \\
OMC~2-3/U  &  0.414 & 1.7$\pm$0.2  & 1.2$\pm$0.2 & 3.0 &    0.09:  &  0.32  &  81  &  100  &  $0.45^{+0.07}_{-0.07}$ \\
ONC~Flank~N/A  & 0.414 & 1.7$\pm$0.2  & 1.2$\pm$0.2 & 3.0 &    0.13  &  0.52  &  80  &  104  &  $0.43^{+0.07}_{-0.07}$ \\
ONC~Flank~S/A  & 0.414 & 1.6$\pm$0.2  & 1.1$\pm$0.1 & 2.8 &    0.09:  &  0.41  &  114  &  122  &  $0.48^{+0.06}_{-0.06}$ \\
RCW~120/B  &  1.35 & 0.8$\pm$0.2  & 0.6$\pm$0.1 & 1.8 &    0.36:  &  1.41  &  41  &  32  &  $0.56^{+0.11}_{-0.11}$ \\
RCW~120/C  &  1.35 & 0.7$\pm$0.4  & 0.5$\pm$0.3 & 1.6 &    0.19:  &  1.07  &  26  &  10  &  $0.72^{+0.16}_{-0.12}$ \\
RCW~120/U  &  1.35 & 1.2$\pm$0.2  & 0.9$\pm$0.2 & 2.5 &    0.15:  &  0.91  &  22  &  40  &  $0.35^{+0.11}_{-0.12}$ \\
Serpens~Main/B  & 0.415 & 0.6$\pm$0.7  & 0.3$\pm$0.8 & 1.2 &    0.10:  &  0.64  &  14  &  10  &  $0.58^{+0.19}_{-0.17}$ \\
Serpens~Main/U  & 0.415 & 2.2$\pm$0.5  & 1.5$\pm$0.5 & 3.6 &    0.09:  &  0.19  &  17  &  19  &  $0.47^{+0.15}_{-0.16}$ \\
Serpens~South/U  & 0.415 & 1.8$\pm$0.8  & 1.7$\pm$0.6 & 4.0 &    0.11:  &  0.31  &  15  &  11  &  $0.58^{+0.19}_{-0.17}$ \\
Sh~2-106/U  &  1.4 & 0.8$\pm$0.4  & 0.6$\pm$0.4 & 1.7 &    0.13:  &  0.60  &  49  &  43  &  $0.53^{+0.10}_{-0.10}$ \\
%\hline
&&&&&&&&\\
\cline{1-10} 
\end{tabular}
\end{minipage}
\end{table}
\clearpage
\newpage

Table~\ref{t:clusters} shows the following information for each of the
69 clusters studied in the current work: region name and cluster
designation; three types of median ages ($Age_{JX}$-Siess00, $Age_{JX}$-MIST, and $Age$-Feiden16M); mass cut ($M_{cut}$) and median stellar mass based on the Siess00 model; the number of disk-bearing ($N_{disk}$) and disk-free ($N_{nodisk}$) YSOs after the imposition of $M_{cut}$; and disk
fraction ($N_{disk}/(N_{disk}+N_{nodisk})$).
In the more complete, on-line version of Table~\ref{t:clusters}, each cluster has four entries, one for each of the four permutations
yielded by using two different YSO classification schemes (catalog classes
versus $\alpha_{IRAC}$-based) and including and excluding probable protostars.

A cluster designation of ``U'' for a SFiNCs cluster indicates the unclustered
component of a region. An $M_{cut}$ value suffixed by a ``:'' indicates that no
YSOs were removed from the sample according to the process described in the
previous subsection, in which case the given value of $M_{cut}$ indicates the
lowest stellar mass in the sample for that cluster. Disk fraction uncertainties
are Wilson confidence intervals for confidence level 0.05 \citep{Wilson1927,
Brown2001}. Other properties of these SFiNCs and MYStIX clusters can be
found in \citet{Getman2018}, \citet{Kuhn2014}, \citet{Kuhn2015a}, and
\citet{Kuhn2015b}.

\section{Results} \label{s:results}

In most of this section, we explore the effects of initial disk fraction (\S\S \ref{s:assume} and \ref{s:noassume}), star-forming environment (\S \ref{s:sfrenv}), and stellar mass (\S\S \ref{s:mvss} and \ref{s:massbins}) on disk longevity, employing age and mass results inferred from a single PMS evolutionary model, namely the Siess00 model. The effect of different PMS evolutionary models on disk longevity estimates is examined in \S \ref{s:pms_models} using the Siess00, MIST, and Feiden16M models.

\begin{figure}
\includegraphics[width=75mm]{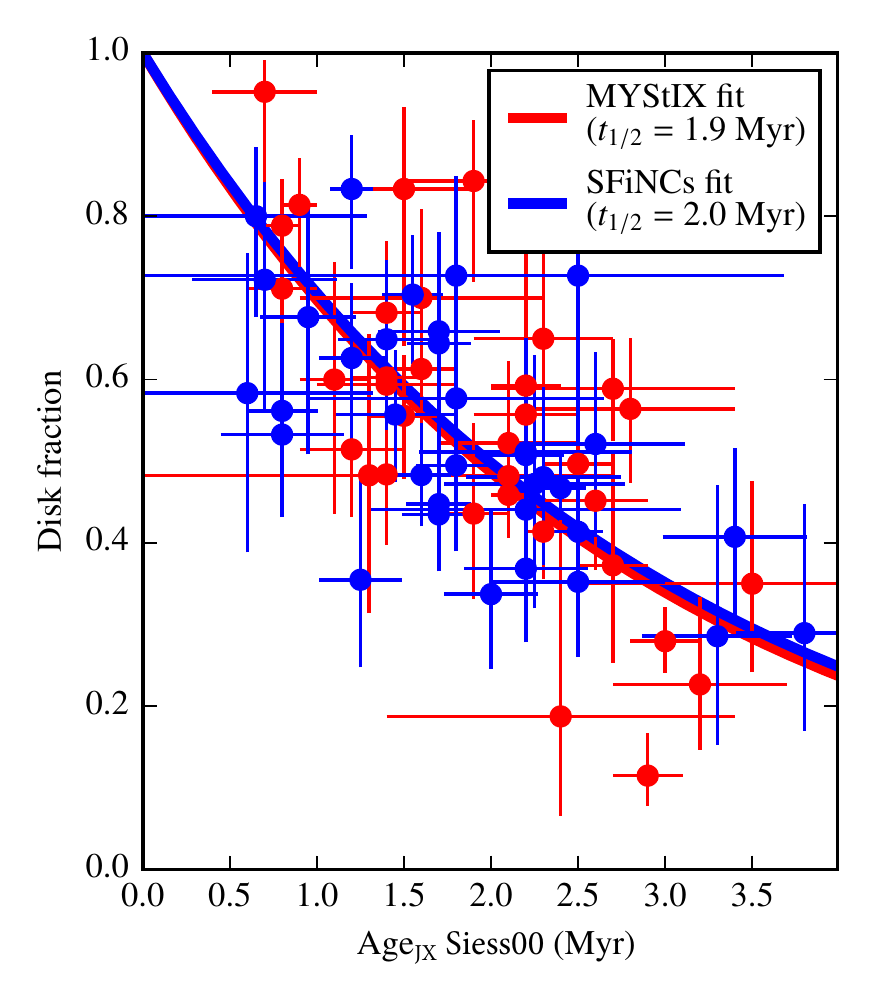}
\caption{Disk fraction versus $Age_{JX}$-Siess00 for 69 MYStIX and SFiNCs clusters (red and blue points, respectively). The current figure exhibits results for one of the four membership permutations, that is, the $\alpha_{IRAC}$-based YSO classification with probable protostars included. Figure panels showing disk fraction as a function of age for the other three permutations (based on catalog YSO classification with/without probable protostars included and based on the SED slope YSO classification when probable protostars are excluded) are provided in the Supplementary Materials. Vertical error bars show Wilson binomial confidence intervals for confidence level 0.05.}
\label{f:DFvsAge}
\end{figure}

\subsection{Disk fraction versus age} \label{s:DFvsAge}
\subsubsection{With assumption of 100\% initial disk fraction} \label{s:assume}

Here we produce a disk evolution plot resembling that of \citet{Haisch2001b} and other researchers. Cluster disk fraction as a function of $Age_{JX}$-Siess00 for 69 MYStIX and
SFiNCs clusters is shown in Fig.~\ref{f:DFvsAge}. The four figure versions (available in the Supplementary Materials) reflect
the four permutations of our analysis methods: two YSO classification schemes
and two rules regarding the inclusion of probable protostars (\S~\ref{s:yso}).
We perform a Gauss--Newton least-squares fit of an exponential function $f_{disk}=f_0 \times e^{-t/\tau}$ (with the mean lifetime $\tau$ as a free parameter) to each
of the 8 datasets (MYStIX and SFiNCs separately for each of the four
membership permutations), assuming initial disk fraction $f_0 = 100\%$. The results are
summarized in Table~\ref{t:longevity}, along with the total number of
disk-bearing and disk-free YSOs included in each analysis, as well as the sum of
the squared residuals (SSR) for each exponential fit. The estimated disk
half-life $t_{1/2}=\tau \times ln(2)$ depends somewhat on the YSO classification scheme used, as
well as on whether probable protostars are included in the sample, however the
variation among these age estimates does not vary beyond their 95\% confidence
intervals (shown in Table~\ref{t:longevity}). The goodness of fit is better for
SFiNCs regions, and is typically slightly better for the analyses using
$\alpha_{IRAC}$-based YSO classifications compared with those using catalog
classifications.
 \begin{table}\scriptsize
\centering
 \begin{minipage}{75mm}
 \caption{Disk longevity estimates ($f_0 = 100$\%).}
 \label{t:longevity}
 \begin{tabular}{@{\vline }c@{ \vline }c@{ \vline }c@{ \vline }c@{ \vline }c@{ \vline }c@{ \vline }c@{ \vline }}
\cline{1-7}
&&&&&&\\ 
%\hline
Project & YSO & Incl. possible & $N_{disk}$ & $N_{nodisk}$ & $t_{1/2}$ & SSR\\
        & classes & protostars? &  &  & (Myr) & \\
(1)&(2)&(3)&(4)&(5)&(6)&(7)\\
%\hline
\cline{1-7}
&&&&&&\\ 
MYStIX  &  Catalog       &  N  &  1329  &  1606  &  $1.7_{-0.2}^{+0.3}$  &  0.90  \\
MYStIX        &  Catalog       &  Y  &  1523  &  1675  &  $1.9_{-0.3}^{+0.3}$  &  0.95  \\
MYStIX        &  $\alpha_{IRAC}$  &  N  &  1924  &  2229  &  $1.8_{-0.2}^{+0.2}$  &  0.53  \\
MYStIX        &  $\alpha_{IRAC}$  &  Y  &  2180  &  2236  &  $1.9_{-0.2}^{+0.3}$  &  0.61  \\
SFiNCs        &  Catalog       &  N  &  1971  &  1778  &  $2.1_{-0.2}^{+0.3}$  &  0.61  \\
SFiNCs        &  Catalog       &  Y  &  2088  &  1761  &  $2.2_{-0.2}^{+0.3}$  &  0.55  \\
SFiNCs        &  $\alpha_{IRAC}$  &  N  &  1924  &  1971  &  $1.9_{-0.2}^{+0.2}$  &  0.59  \\
SFiNCs        &  $\alpha_{IRAC}$  &  Y  &  2062  &  1988  &  $2.0_{-0.2}^{+0.2}$  &  0.54  \\
%\hline
&&&&&&\\ 
\cline{1-7} 
\end{tabular}
\end{minipage}
\end{table}

\subsubsection{Without assumption of 100\% initial disk fraction} \label{s:noassume}

We explore the possibility that the zero-age disk fraction $f_0$ is less
than 100\%. That is, some stars would be born without dusty circumstellar disks, or lose these disks very rapidly ($<0.5$~Myr). As in the previous subsection, we fit exponential functions, this
time with $f_0$ as an additional parameter, using the adaptive nonlinear
least-squares algorithm of \citet{Dennis1981}. Results are shown in
Table~\ref{t:secondlongevity}. For MYStIX clusters, the exponential
half-lives do not differ significantly from those reported in
Table~\ref{t:longevity}, though the uncertainties become considerably larger;
the initial disk fractions remain consistent with 100\%. For SFiNCs
clusters, the estimated half-lives are longer than those shown in
Table~\ref{t:longevity}, but no more than $2 \sigma$.
In the two SFiNCs analyses where probable protostars are excluded, the upper 95\%
confidence intervals for $f_0$ only reach the low-90\% range, which leaves room for
the possibility that not all YSOs begin with a hot inner dust
disk.
\begin{table}\scriptsize
\centering
 \begin{minipage}{75mm}
 \caption{Disk longevity and initial disk fraction estimates ($f_0 < 100$\%).}
 \label{t:secondlongevity}
 \begin{tabular}{@{\vline }c@{ \vline }c@{ \vline }c@{ \vline }c@{ \vline }c@{ \vline }c@{ \vline }}
\cline{1-6}
&&&&&\\ 
%\hline
Project & YSO & Incl. possible & $t_{1/2}$ & $f_0$ &  SSR\\
        & classes & protostars? & (Myr) & & \\
(1)&(2)&(3)&(4)&(5)&(6)\\
%\hline
\cline{1-6}
&&&&&\\ 
MYStIX  &  Catalog       &  N  &  $2.2_{-0.7}^{+2.0}$  &  $0.84_{-0.21}^{+0.16}$  &  0.79  \\
MYStIX        &  Catalog       &  Y  &  $2.3_{-0.7}^{+2.2}$  &  $0.87_{-0.22}^{+0.13}$  &  0.88  \\
MYStIX        &  $\alpha_{IRAC}$  &  N  &  $1.8_{-0.2}^{+0.6}$  &  $1.00_{-0.18}^{+0.00}$  &  0.53  \\
MYStIX        &  $\alpha_{IRAC}$  &  Y  &  $1.9_{-0.2}^{+0.5}$  &  $1.00_{-0.13}^{+0.00}$  &  0.61  \\
SFiNCs        &  Catalog       &  N  &  $3.5_{-1.2}^{+3.9}$  &  $0.77_{-0.14}^{+0.16}$  &  0.43  \\
SFiNCs        &  Catalog       &  Y  &  $3.2_{-1.0}^{+2.7}$  &  $0.83_{-0.14}^{+0.16}$  &  0.43  \\
SFiNCs        &  $\alpha_{IRAC}$  &  N  &  $3.2_{-1.0}^{+2.9}$  &  $0.75_{-0.14}^{+0.16}$  &  0.43  \\
SFiNCs        &  $\alpha_{IRAC}$  &  Y  &  $2.9_{-0.9}^{+2.2}$  &  $0.81_{-0.15}^{+0.17}$  &  0.44\\
%\hline
&&&&&\\ 
\cline{1-6} 
\end{tabular}
\end{minipage}
\end{table}

The large uncertainties on the disk half-life estimates shown in
Table~\ref{t:secondlongevity} stem from not having clusters in our sample that are both younger than 0.5~Myr and
significantly older than the estimated $e$-folding times. Future works that examine disk longevity over a larger
age range and have much richer homogeneous cluster samples should omit the assumption of 100\% disk fraction at zero age, as well as explore other parametrizations
of disk longevity, given that there is no physical basis for assuming an exponential decay.

\subsubsection{Dependence on star-forming environment}\label{s:sfrenv}

Using the MYStIX and SFiNCs datasets, we attempt to test whether disks in the richer star-forming environments targeted by the MYStIX
survey evolve differently from disks in the more sparse environments seen in SFiNCs fields. In particular, MYStIX clusters are often dominated by multiple O stars while SFiNCs clusters are generally dominated by a single massive star, typically a late-O or early-B.

Previous observations and theory suggest no strong effects by OB photoevaporation and/or dynamical interactions of cluster members on the inner parts of circumstellar disks. For instance, the results of \citet{Richert2015} suggest that external
photoevaporation by OB stars does not affect the presence of infrared excess
\citep[nor is it likely to based on theory, which does not predict significant
truncation closer than ${\sim}100$~AU to a disk's host star; e.g.,][]{Anderson2013}.
Meanwhile, theory regarding disk truncation due to dynamical encounters among
YSOs does not predict disk depletion much closer to the star than
${\sim}100$~AU, and certainly not within the few-AU orbital radii associated
with near-/mid-infrared excess \citep{PortegiesZwart2016, Vincke2016}. 

We find that the estimated values of $t_{1/2}$ and $f_0$
shown in Tables~\ref{t:longevity} and \ref{t:secondlongevity} do not differ beyond the expected statistical error
between MYStIX and SFiNCs. Thus our data provide no evidence for distinctive dissipation timescales of disks in different star-forming environments.

\subsection{Effect of PMS evolutionary models on disk longevity} \label{s:pms_models}

Here we investigate the effect of uncertainty in our knowledge of PMS evolution on disk longevity. We study disk fraction as a function of age using three different sets of cluster ages: the traditional $Age_{JX}$-Siess00, the new $Age_{JX}$-MIST with improved microphysics, and the new $Age$-Feiden16M that include ``radius inflation'' due to magnetic fields (\S \ref{s:ages}). This analysis is applied to the combined MYStIX+SFiNCs sample of 69 clusters using four different membership permutations (Table \ref{t:longevity_vs_model}). Using the nonlinear weighted Gauss--Newton least-squares method, the datasets are fit with an exponential function $f_{disk}=f_0 \times e^{-t/\tau}$ with the mean lifetime $\tau$ as a free parameter and the initial disk fraction ($f_0$) fixed at 100\%.

\begin{figure}
\includegraphics[width=170mm]{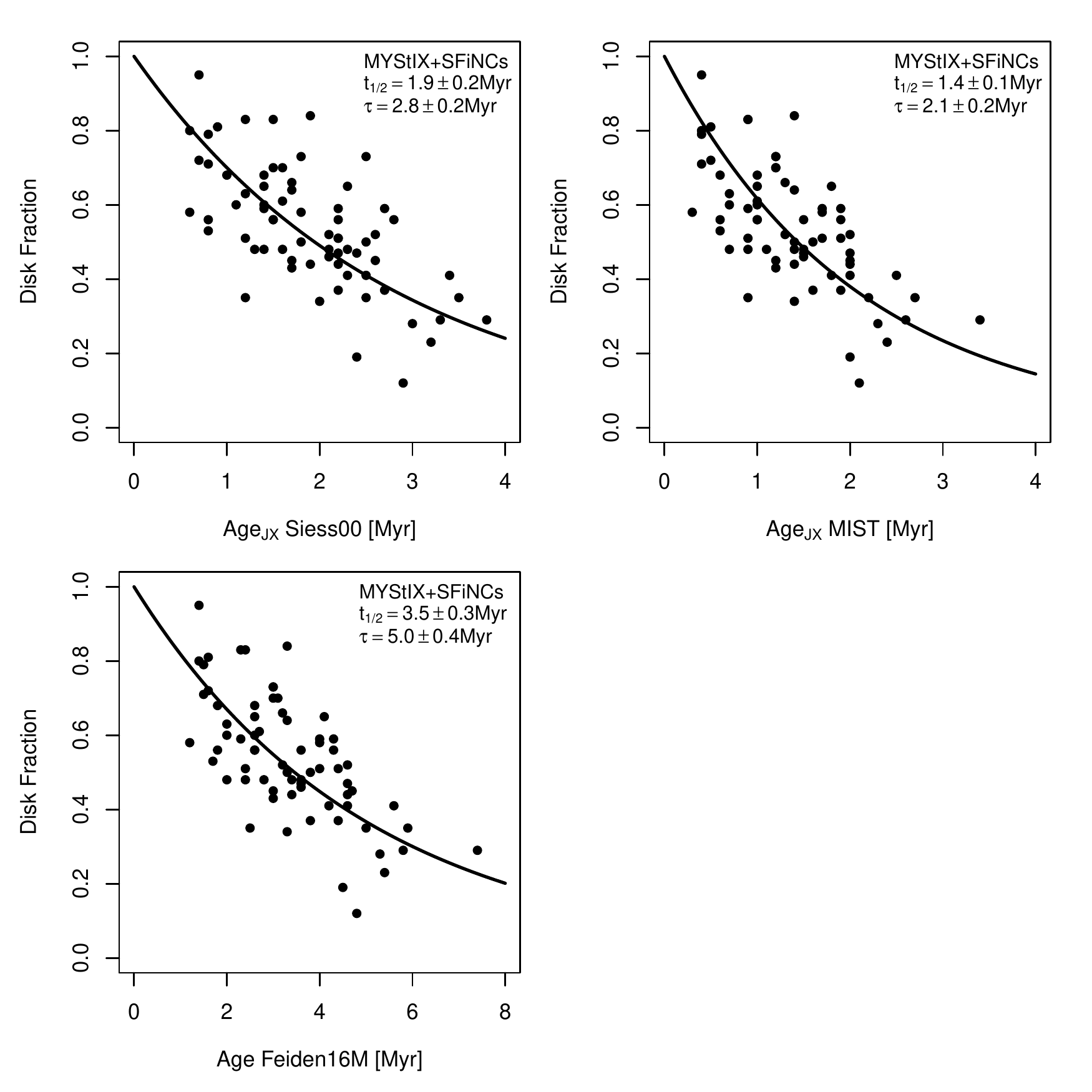}
\caption{Disk fraction as a function of age using 3 types of cluster ages ($Age_{JX}$-Siess00, $Age_{JX}$-MIST, and $Age$-Feiden16M). The underlying dataset (black points) is the combined MYStIX+SFiNCs sample of 69 clusters. The current figure exhibits results for one of the four membership permutations, that is, the $\alpha_{IRAC}$-based YSO classification with probable protostars included. Figure panels showing disk fraction as a function of age for the other three permutations are provided in the Supplementary Materials. The black lines represent the best model fits. The figure legends provide disk dissipation timescales resulting from the fits to an exponential model.}
\label{f:df_age_threemodels}
\end{figure}

Fig.~\ref{f:df_age_threemodels} and Table~\ref{t:longevity_vs_model} show that even though the goodness of fit is better for the cases with ``possible protostars included'', the variations in $t_{1/2}$ due to different membership permutations are small ($<15$\% of a value) and are comparable to the statistical errors. 
\begin{table}\scriptsize
\centering
 \begin{minipage}{75mm}
 \caption{Disk longevity from the use of different PMS evolutionary models.}
 \label{t:longevity_vs_model}
 \begin{tabular}{@{\vline }c@{ \vline }c@{ \vline }c@{ \vline }c@{ \vline }c@{ \vline }c@{ \vline }}
\cline{1-6}
&&&&&\\ 
%\hline
Project & PMS & YSO & Incl. & $t_{1/2}$ & SSR\\
        & evolutionary & classes & possible & (Myr) &\\
        & model     &            & protostars? &  &\\
(1)&(2)&(3)&(4)&(5)&(6)\\
%\hline
\cline{1-6}
&&&&&\\ 
MYStIX+SFiNCs & Siess00      & Catalog & N       & $1.9 \pm 0.2$   &  2.23 \\
MYStIX+SFiNCs       & Siess00      & Catalog & Y       & $2.0 \pm 0.2$   &  1.52 \\
MYStIX+SFiNCs       & Siess00      &$\alpha_{IRAC}$ & N   & $1.8 \pm 0.1$   &  1.64 \\
MYStIX+SFiNCs       & Siess00      & $\alpha_{IRAC}$ & Y  & $1.9 \pm 0.2$   &  1.15 \\
MYStIX+SFiNCs       & MIST         & Catalog & N       & $1.4 \pm 0.1$   &  2.44 \\
MYStIX+SFiNCs       & MIST         & Catalog & Y       & $1.5 \pm 0.2$   &  1.66 \\
MYStIX+SFiNCs       & MIST         &$\alpha_{IRAC}$ & N   & $1.3 \pm 0.1$   &  1.67 \\
MYStIX+SFiNCs       & MIST         & $\alpha_{IRAC}$ & Y  & $1.4 \pm 0.1$   &  1.17 \\
MYStIX+SFiNCs       & Feiden16M    & Catalog & N       & $3.4 \pm 0.3$   &  2.06 \\
MYStIX+SFiNCs       & Feiden16M    & Catalog & Y       & $3.6 \pm 0.4$   &  1.43 \\
MYStIX+SFiNCs       & Feiden16M    &$\alpha_{IRAC}$ & N   & $3.2 \pm 0.2$   &  1.46 \\
MYStIX+SFiNCs       & Feiden16M    & $\alpha_{IRAC}$ & Y  & $3.5 \pm 0.3$   &  1.04 \\
%\hline
&&&&&\\ 
\cline{1-6} 
\end{tabular}
\end{minipage}
\end{table}

In contrast, the choice of different PMS models has a much stronger effect (70\%-170\%) on estimated $t_{1/2}$. The magnetic PMS models of \citet{Feiden2016} lead to significantly longer disk dissipation timescales ($t_{1/2,Feiden16M} \sim 3.5$~Myr) compared to those of the non-magnetic models ($t_{1/2,MIST,Siess00} \sim 1.3-2.0$~Myr). Clearly the choice of stellar evolutionary models, especially when magnetic fields are included, have a major effect on the inferred half-life of inner disks.

\clearpage
\newpage

\clearpage
\newpage

\subsection{The role of stellar mass} \label{s:masses}

Several previous works have indicated that disks survive longer around lower mass stars than around higher mass stars \citep{Haisch2001a, Carpenter2006, Luhman2012, Ribas2015}.

In Section~\ref{s:mvss}, we compare the properties of MYStIX and SFiNCs
regions in order to determine differences in disk fraction due to the greater prevalence of higher mass stars in MYStIX than SFiNCs samples.
In Section~\ref{s:massbins}, we analyze the combined MYStIX and
SFiNCs samples by binning YSOs according to age ($Age_{JX}$-Siess00) and stellar mass rather than
by cluster.
\begin{figure}
\includegraphics[width=75mm]{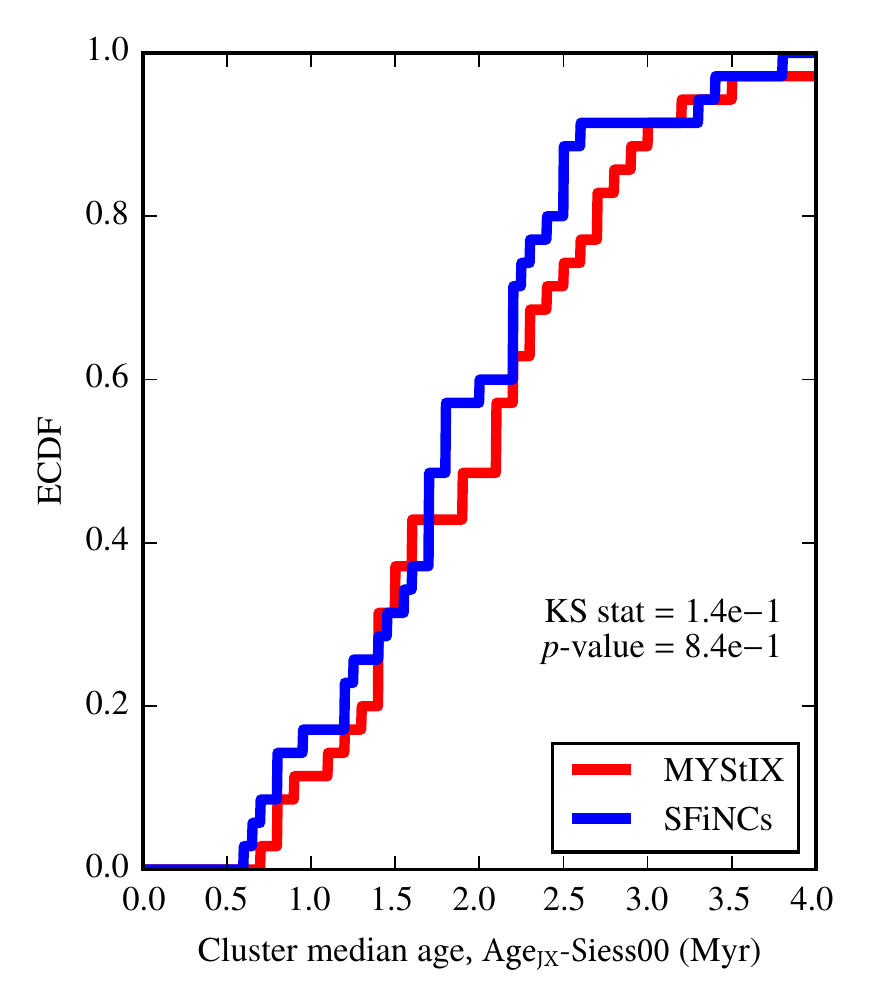}
\caption{ECDF of median cluster ages ($Age_{JX}$-Siess00) for MYStIX (red) and SFiNCs (blue).}
\label{f:ageECDF}
\end{figure}

\subsubsection{Comparing MYStIX and SFiNCs clusters} \label{s:mvss}

We perform several comparisons of MYStIX and SFiNCs cluster properties.
First, in Fig.~\ref{f:ageECDF}, we compare ECDFs of median cluster ages (based on the Siess00 model) for
MYStIX and SFiNCs. The ages of the MYStIX and SFiNCs cluster samples are similar. 

\begin{figure}
\begin{tabular}{cc}
%\begin{minipage}{100mm}
\includegraphics[width=0.45\textwidth]{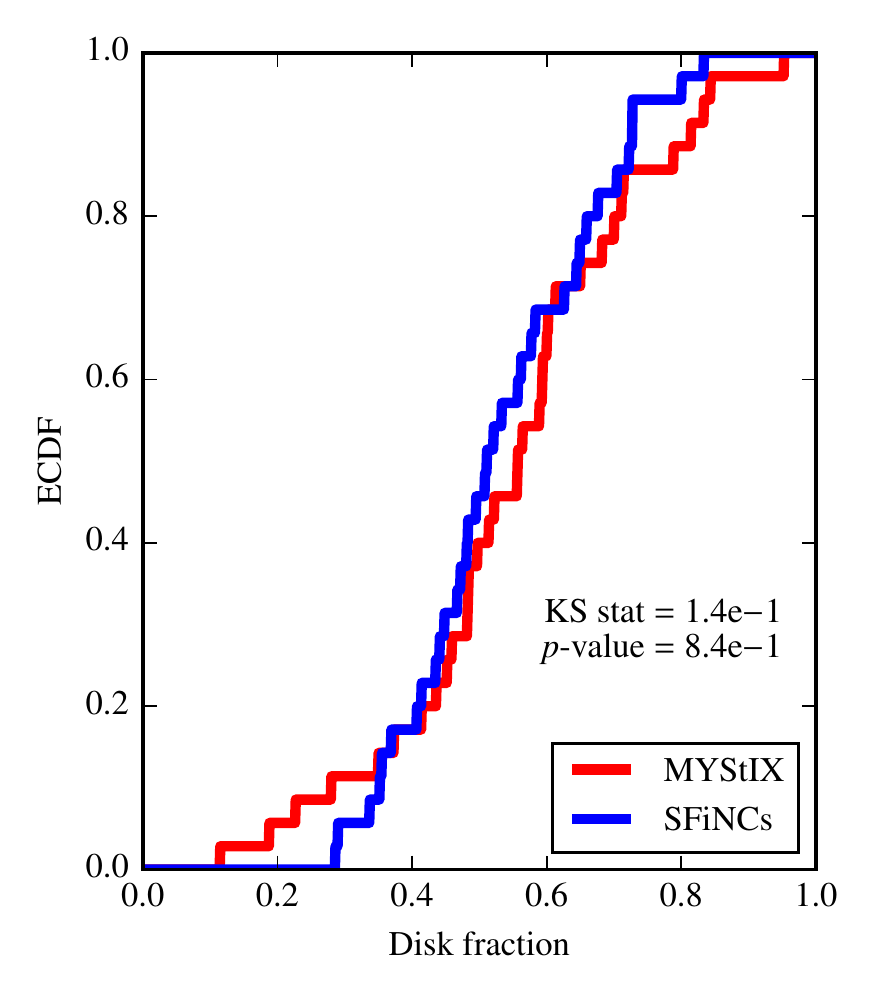}
\includegraphics[width=0.45\textwidth]{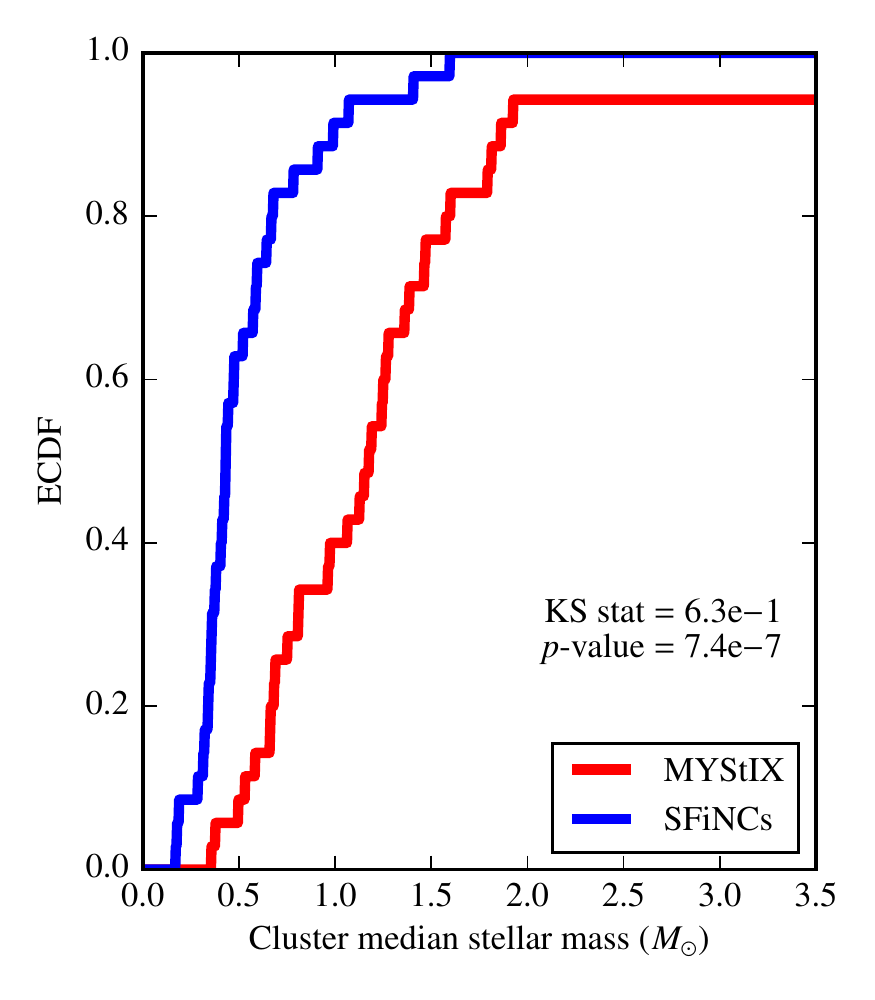}
\end{tabular}%
\begin{tabular}{cc}
\includegraphics[width=0.45\textwidth]{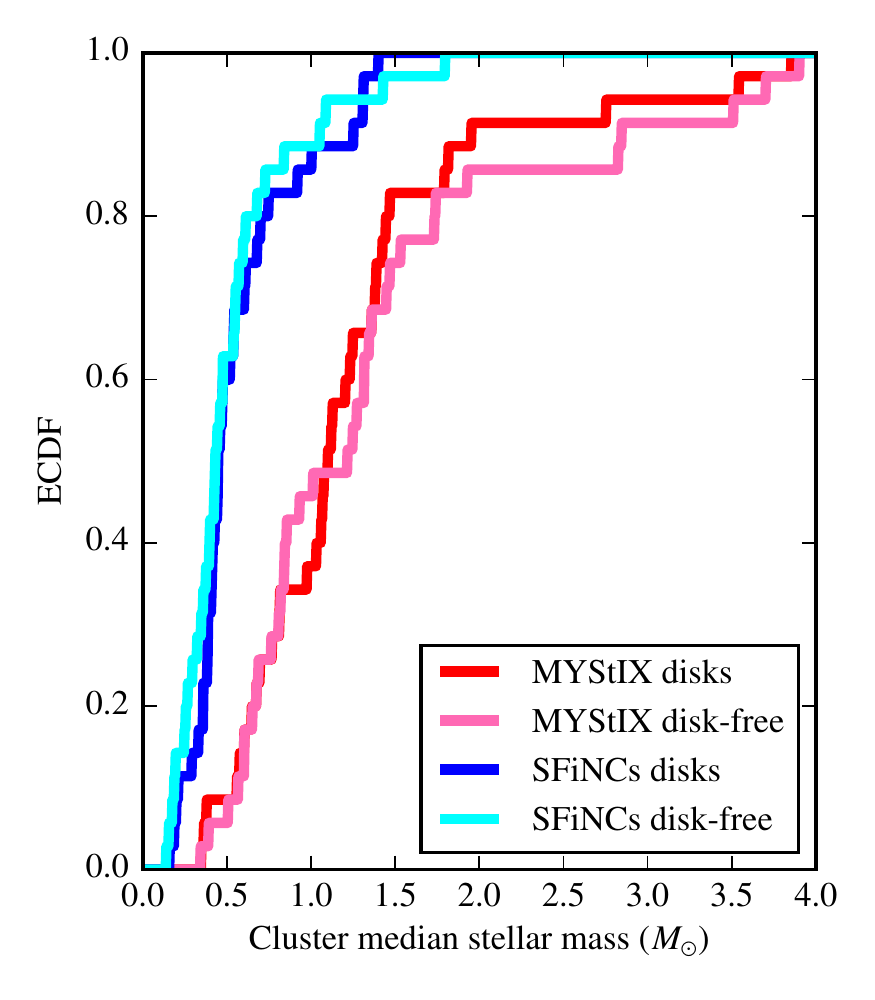}
\includegraphics[width=0.45\textwidth]{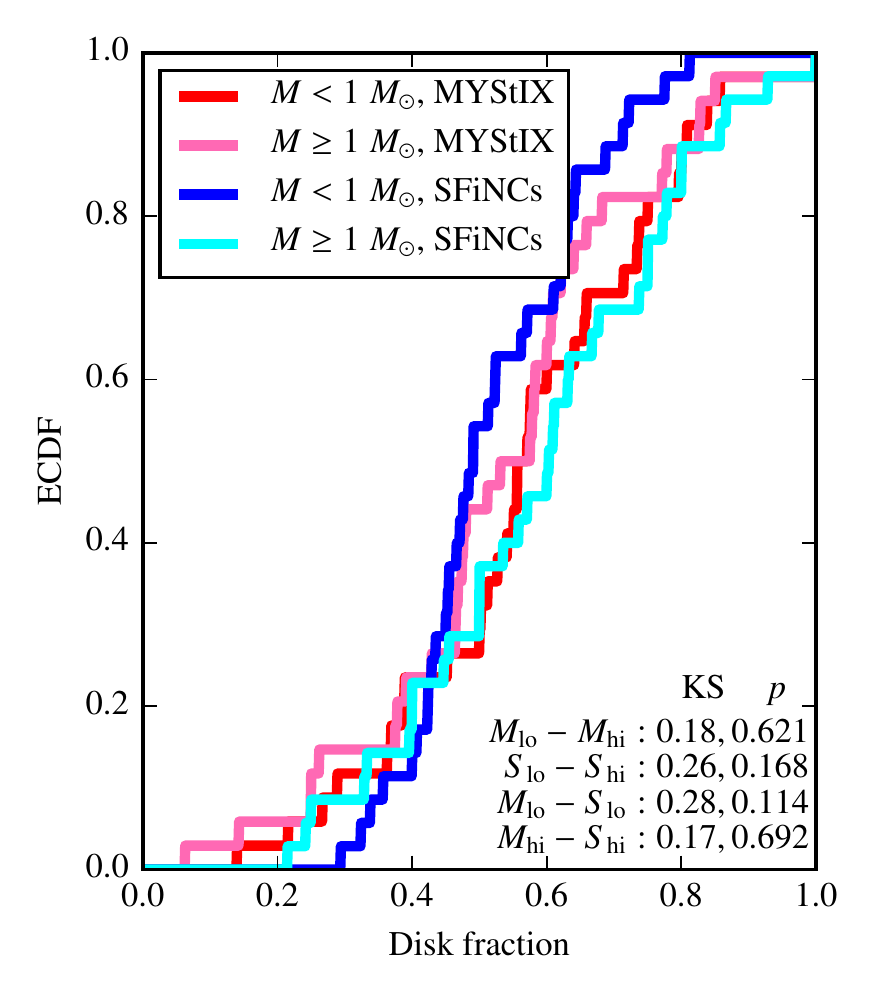}
\end{tabular}
\caption{Cumulative distributions of cluster disk fraction (upper left panel),
cluster median stellar mass (based on the Siess00 model; upper right panel), cluster median stellar mass
separately for disk-bearing and disk-free YSOs (lower left panel), and disk fraction separately for low-mass ($M<1$~M$_{\odot}$) and high-mass ($M\ge1$~M$_{\odot}$) YSOs (lower right panel), separately for MYStIX and SFiNCs. The current figure shows a membership case with YSOs classified based on $\alpha_{IRAC}$ value
and probable protostars included ($\alpha_{IRAC}>0$). Figure panels showing cumulative distributions of disk fraction and median mass for the other three membership permutations are provided in the Supplementary Materials.}
\label{f:ECDFs_sedslope_incl}
%\end{minipage}
\end{figure}

Fig.~\ref{f:ECDFs_sedslope_incl} shows cluster disk fraction (across entire mass range and mass-stratified), median stellar mass (based on the Siess00 model), and median stellar mass by YSO type for the following analysis schemes, in order: catalog YSO classifications and
excluding probable protostars; catalog YSO classifications and
including probable protostars; $\alpha_{IRAC}$-based YSO classifications and
excluding probable protostars; and $\alpha_{IRAC}$-based YSO classifications and including
probable protostars.

The upper left panels of Fig.~\ref{f:ECDFs_sedslope_incl} show only
modest differences in the distributions of disk fractions between MYStIX and
SFiNCs clusters, which appear to stem largely from the presence of a handful
of MYStIX clusters with low disk fractions. This is consistent with the
similarity of the estimates of exponential disk half-life shown in
Table~\ref{t:longevity}, and does not provide strong evidence for differential
disk longevity between sparse (SFiNCs) and rich (MYStIX) star-forming
regions.

The upper right panels of Fig.~\ref{f:ECDFs_sedslope_incl} show ECDFs of
median cluster stellar mass for MYStIX and SFiNCs (YSO mass derivations
are discussed in \S~\ref{s:clustersummary}). MYStIX clusters have higher
median masses for two reasons. First, MYStIX clusters are more distant,
therefore a given magnitude limit in a given near-infrared band will translate
into a higher stellar mass. Second, MYStIX clusters are richer and therefore
physically contain more of the rare high-mass stars.

The lower left panels of Fig.~\ref{f:ECDFs_sedslope_incl} show ECDFs of
subcluster median stellar mass separately for disk-bearing and disk-free YSOs,
and separately for MYStIX and SFiNCs (median masses are calculated after
$M_{cut}$ has been applied). The disk-bearing and disk-free
distributions are quite close to each other within both MYStIX and SFiNCs
while the distributions between the two projects are quite different.

The lower right panels of Fig.~\ref{f:ECDFs_sedslope_incl} show ECDFs of disk fraction stratified by mass, separately for SFiNCs and MYStIX clusters. To mitigate the small number statistics issue (some clusters have only a few stars with $M>2$~M$_{\odot}$) the mass cutoff is chosen here as $1$~M$_{\odot}$. The figure shows no differences in the distributions of disk fractions between higher- and lower-mass stars when using the $\alpha_{IRAC}$-based YSO classifications. Only modest differences in the distributions of disk fractions for SFiNCs stars are present when using the Catalog YSO classifications; with a hint of possible age dependence --- the SFiNCs clusters with higher disk fractions for more massive stars are, on average, slightly younger than the SFiNCs clusters with lower disk fractions for more massive stars. But this age difference is not statistically significant.

To summarize, for the same age range, MYStIX clusters have higher median masses (upper right and lower left panels of Fig.~\ref{f:ECDFs_sedslope_incl}) than SFiNCs clusters. Any mass dependence of disk longevity thus should be apparent in the upper left and lower right panels of Fig.~\ref{f:ECDFs_sedslope_incl}. The similarities of disk fractions between SFiNCs and MYStIX clusters when using the $\alpha_{IRAC}$-based YSO classifications and only modest differences for some cluster sub-samples when using the Catalog YSO classifications show that disk fraction is not strongly dependent on stellar mass. We explore this result in more detail in the following subsection.
  
\newpage
\clearpage
\newpage

\subsubsection{Disk longevity as a function of host star mass} \label{s:massbins}

For the stellar population in the relatively old Upper~Sco association \citep[$t \sim 11$~Myr;][]{Pecaut2012}, \citet{Carpenter2006, Luhman2012} find that the more massive stars (M$>1.2$~M$_{\odot}$) have a lower fraction of inner primordial disks. For the younger star-forming regions, NGC~1333 and IC~348 ($t \la 3$~Myr), \citet{Luhman2016} find no signs of disk fraction variations within the spectral type range from L-type to B-type. Similarly, \citet{Kennedy2009} find no significant differences in disk fraction between the M$<1.5$~M$_{\odot}$ and M$>1.5$~M$_{\odot}$ members of the younger, $t \la 3$~Myr, stellar populations in Taurus, Cha~I, and IC~348, but they provide observational evidence of stellar mass-dependent disk dispersal for older ($t > 3$~Myr) populations. In contrast, \citet{Ribas2015} suggest that the differences in the primordial disk fraction between the higher-mass ($>2$~M$_{\odot}$) and lower-mass ($<2$~M$_{\odot}$) stars are present even in younger ($t \la 3$~Myr) regions, such as in the combined sample of Cha~I, Cha~II, CrA, Lupus, NGC~1333, $\sigma$~Ori, Serpens, and Taurus.

Physically, this implies either an increased disk
depletion rate due to accretion, photoevaporation, and planet(esimal) formation
for higher-mass host stars, or that initial disk fraction decreases with
increasing stellar mass. In order to explore this question using the MYStIX
and SFiNCs datasets, we place the YSOs that informed Fig.~\ref{f:DFvsAge}
into four stellar mass bins and four age bins; in other words, YSOs are now
associated by age and mass, not by cluster. In this analysis, cluster ages and masses are derived based on the Siess00 model.

\begin{figure}
\includegraphics[width=75mm]{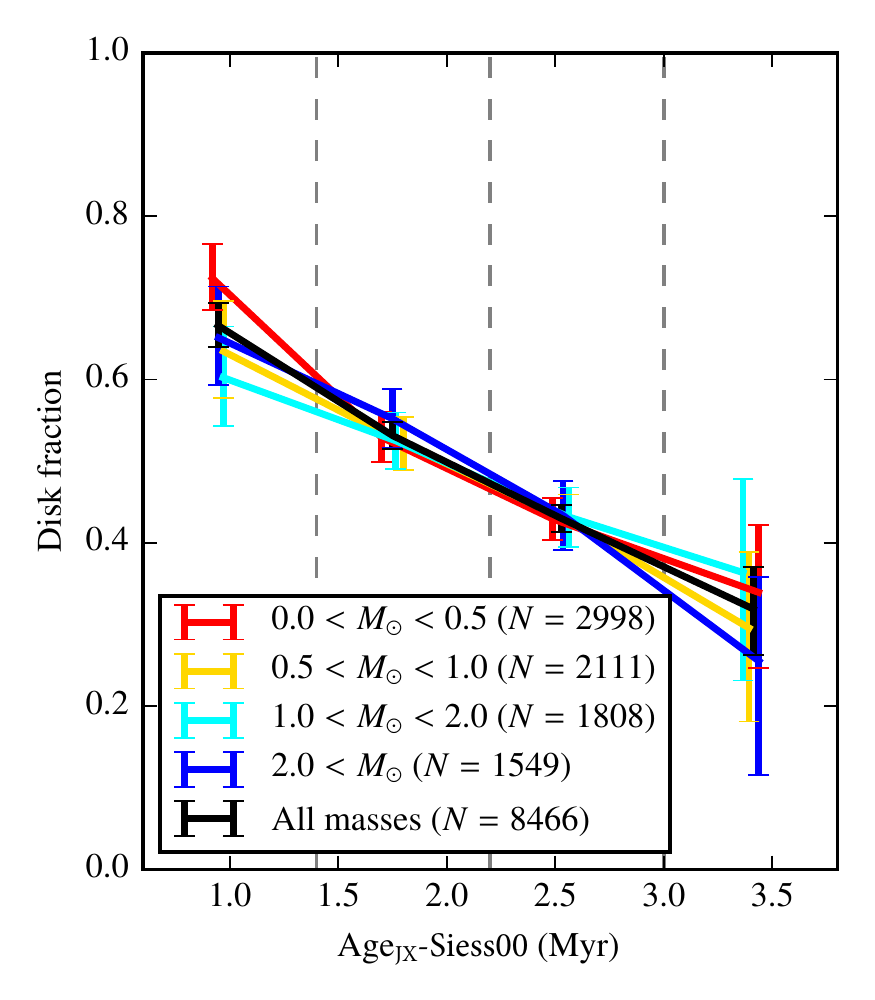}
\caption{Disk fraction as a function of age ($Age_{JX}$-Siess00) for four bins of stellar mass,
combining data across all 69 MYStIX and SFiNCs clusters. The current figure exhibits results for one of the four membership permutations, that is, the $\alpha_{IRAC}$-based YSO classification with probable protostars included. Figure panels showing disk fraction as a function of age for the other three permutations (based on catalog YSO classification with/without probable protostars included and based on the SED slope YSO classification when probable protostars are excluded) are provided in the Supplementary Materials. Vertical error bars are Wilson binomial confidence intervals for significance level 0.05. Dashed lines indicate age bin boundaries.}
\label{f:massbins}
\end{figure}

Disk fraction versus age for each of the four mass bins is shown in
Fig.~\ref{f:massbins}; the sample size within each mass bin is shown in the
legend for each panel; with star numbers varying within the 1,300--2,800 range. The figure shows
no statistically significant trends in disk fraction with stellar mass for young
($<4$~Myr; assuming the Siess00 age scale) clusters.

Surveys exclusively using infrared data to detect and classify YSOs may be at
risk of underestimating the number of disk-free YSOs and thereby overestimating
disk fraction (and subsequently disk longevity) among low-mass stars. By
applying $M_{cut}$ (\S~\ref{s:sensitivity}), we ensure similar sensitivity to
disk-bearing and disk-free YSOs.

However, our survey may have a bias against disk-free intermediate-mass stars because it relies on X-ray selection. It is
well known that the X-ray sensitivity diminishes towards mid-B and mid-A type
stars \citep[e.g.,][]{Stelzer2009, Gudel2009, Drake2014}. Such an X-ray
sensitivity bias towards A- and B-type stars might lead to overestimation of
disk fractions for our $>2$~M$_{\odot}$ stellar sample in Fig.~\ref{f:massbins}, and
thus prevent us from detecting the effect of lower disk fractions in higher mass
stars reported in a number of previous disk fraction studies.

On the other hand, we consider our non-detection of such a trend to be consistent with the recent findings of \citet{Luhman2016} for the young ($t \la 3$~Myr on the Siess00 age scale) IC~348 and NGC~1333 star-forming regions, which show no statistically significant variations in disk fraction for objects with spectral types between L0 and late B, although their sample of stellar members with spectral types earlier than K6 ($M \ga 1$~M$_{\odot}$) is relatively small. Meanwhile, for the much older population in the Upper~Sco association \citep[$t \sim 11$~Myr;][]{Pecaut2012}, \citet{Luhman2012} report a trend of a decreasing disk fraction from M-type to B--G type stars. These results may point to an increased disk depletion rate in higher-mass stars throughout their disk evolution.

In Fig.~\ref{f:massbins}, for the cases of catalog-based YSO classification with and without probable protostars, we see diminished disk fractions for young (youngest two age bins), intermediate-mass (1--2~M$_{\odot}$) stars. To
explore this, we first redo the analysis shown in Fig.~\ref{f:massbins}
separately for MYStIX and SFiNCs (results not shown); the effect seems to
emerge from the MYStIX sample alone. We also find that the increase in disk
fraction going from the catalog-based to the SED-based classifications comes from a gain in
disk-bearing YSOs, not a loss of disk-free ones. Given the high nebulosity in
many MYStIX regions, it is reasonable to suspect that contamination due to
emission from polycyclic aromatic hydrocarbons (PAHs) reduces the sensitivity of
YSO classification based on SED models toward disk-bearing YSOs. Indeed, we find
that if we reproduce the analysis shown in Fig.~\ref{f:massbins} while
excluding the three most heavily contaminated MYStIX clusters identified by \citet{Richert2015} based on 8~{\micron} background levels,
the effect in question disappears. This would seem to explain the large
differences in $M_{cut}$ between the catalog YSO classifications and SED
slope-based YSO classifications seen in Table~\ref{t:clusters}, such as for
NGC~6357~B, where the SED slope-based classifications achieve stellar mass
completeness down to ${\sim}0.1$~M$_{\odot}$, as opposed to ${\sim}0.6$~M$_{\odot}$ for the
catalog classifications. X-ray measurements do not suffer from this problem,
therefore the number of disk-free YSOs in the young-age, intermediate-mass
regime in question does not change significantly between the catalog-based to SED-based classifications
(Fig.~\ref{f:massbins}).

\citet{Richert2015} find that in several distant, rich, OB-dominated MYStIX
regions, the large point spread functions of early O stars in {\it Spitzer}/IRAC
bands diminish sensitivity toward infrared excess from disk-bearing YSOs
(whereas disk-free objects are still detected by X-rays). To ensure that the
lower disk fraction for high-mass YSOs seen in Fig.~\ref{f:massbins} is not
due to this effect, we perform the same analysis again, but now excluding
OB-dominated regions identified as problematic by \citet{Richert2015}, namely
M~17 and NGC~6357. The results are not affected, apart from the effect discussed
in the previous paragraph.

\section{Conclusions} \label{s:conclusions}
\subsection{Summary of results}

In this work, we have studied circumstellar disk longevity in 69 young
stellar clusters by combining X-ray and infrared data and studying
cluster disk fraction as a function of age. We have applied
homogeneously-derived cluster ages (based on the $Age_{JX}$ method) and carefully
accounted for the relative sensitivity to different clusters in the
infrared and X-ray bands. The SFiNCs and MYStIX samples collectively
exceed previous cluster samples by more than a factor of three. 

Our data show that disk longevity estimates are strongly sensitive to the choice of PMS evolutionary model, but are not so sensititive to YSO classification scheme, initial disk fraction, stellar mass, and star-forming environment.

Our analysis has yielded IRAC half-lives of $t_{1/2} \sim 1.3-2$~Myr based on the non-magnetic Siess00 and MIST models, but much longer half-lives of $t_{1/2} \sim 3.5$~Myr based on the magnetic Feiden16M model. According to the relationship $\tau = t_{1/2} / ln(2)$, these half-lives $t_{1/2}$ translate into mean lifetimes of $\tau \sim 1.9-2.9$~Myr based on the non-magnetic Siess00 and MIST models, and $\tau \sim 5.0$~Myr based on the magnetic Feiden16M model.

Half-life estimates change only somewhat when the initial disk fraction
is allowed to vary below 100\%, however the constraints on initial disk fraction
and especially half-life are weak due to the limited age range of our sample.

We find no statistically significant evidence that disk fraction varies with
stellar mass within the first few Myr of life. However, this result may be inaccurate for MYStIX and SFiNCs stars more massive than $2$~M$_{\odot}$ due to reduced X-ray sensitivity towards mid-B and mid-A type stars.

Our data do not provide clear evidence that disk longevity depends on the surrounding star-forming
environment.

\subsection{Comparison with previous works}

\begin{figure}
\includegraphics[width=75mm]{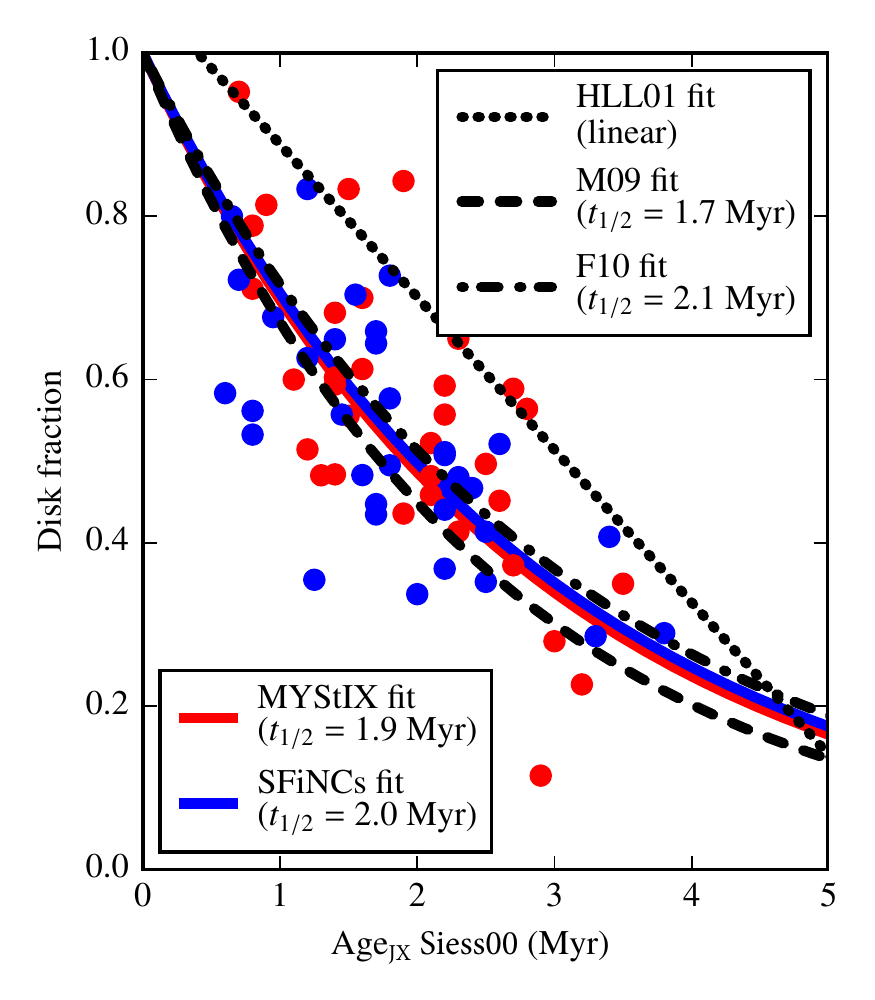}
\caption{Disk fraction versus age (based on the Siess00 models) for 69 MYStIX and SFiNCs clusters, along with results of \citet[][HLL01]{Haisch2001b}, \citet[][M09]{Mamajek2009}, and \citet[][F10]{Fedele2010}. The MYStIX and SFiNCs fits are shown for the case of an SED slope-based YSO classification with probable protostars ($\alpha_{IRAC}>0.0$) included.}
\label{f:DFvsAge1}
\end{figure}

Our finding of a $t_{1/2} {\sim} 2$~Myr half-life (mean lifetime $\tau \sim 2.9$~Myr) for disks based on the Siess00 age scale agrees
closely with those of several previous works \citep{Mamajek2009, Fedele2010, Ribas2014}, which generally included heterogeneous compilations of cluster ages based on old-generation PMS evolutionary models. Fig.~\ref{f:DFvsAge1} shows a comparison of our results with those of \citet{Haisch2001b}, \citet{Mamajek2009}, and
\citet{Fedele2010}. Our finding of a $t_{1/2} {\sim} 3.5$~Myr half-life (mean lifetime $\tau \sim 5.0$~Myr) based on the Feiden16M age scale is consistent with that of the recent work of \citet{Pecaut2016}. Detailed comparison with these studies follows.
\begin{enumerate}

\item {\bf \citet{Haisch2001b}.} Fig.~\ref{f:DFvsAge1} shows that our disk fractions (based on the Siess00 model) are systematically lower than those of
\citet[][HLL01]{Haisch2001b}. One reason is differing disk
classifications. The cluster IC~348 is found in both the SFiNCs and HLL01
samples, making it a useful example for exploring this possibility. In IC~348,
${\sim}$40\% of objects classified as disk-bearing by HLL01 are classified as disk-free in
the SFiNCs catalog. This indicates that the $JHKL$-based color--color
diagram approach to identifying disk-bearing YSOs yields different results from
those provided by longer-wavelength (3--8~$\micron$) data. A similar disparity in the disk classifications is provided by HLL01 and
\citet{Lada2006}, the latter of which makes use of {\it Spitzer} photometry;
approximately a third of the YSOs shared between the HLL01 and \citet{Lada2006}
catalogs have disparate classifications. \citet{Lada2006} notice that the $K -$[3.6] colors of the stars in their work ``appear to be bluer than the $K-L$ colors of Haisch et al.'', for unclear reasons. The SFiNCs {\it Spitzer}-IRAC photometric measurements are consistent with those of Lada et al. \citep[see Figure A3 in][]{Getman2017}. It is worth noting that our estimated
disk fraction for IC~348 agrees closely with the estimate of \citet{Luhman2016}.
The level of 3.6~$\micron$ variability estimated by \citet{Flaherty2013} is small
(standard deviations of order hundredths of a magnitude), and is therefore
unlikely to lead to significant misclassification of YSOs. 

A second possible explanation
for the systematically higher disk fractions reported by HLL01 is related to their imposition 
of different mass limits for disk-bearing and disk-free objects as we discuss in \S \ref{s:sensitivity}. HLL01 use the analysis of \citet{Haisch2001a} to derive the disk fraction of
IC~348. \citet{Haisch2001a} claim to be complete down to $L\sim12$ for
disk-bearing and disk-free objects. However, in $J$ and $H$ bands, which
trace the stellar photosphere and therefore serve as proxies for stellar mass,
the sample of \citet{Haisch2001a} is substantially deeper for disk-bearing
objects. $L$-band measurements are sensitive to the presence of a disk,
therefore the imposition of a similar limit for disk-bearing and disk-free
objects in the $L$ band will be biased toward finding disks, leading to a
significant overestimation of disk fraction. Excluding sources with
$H>12$ for the \citet{Haisch2001a} catalog in IC~348 lowers the disk
fraction by a factor of ${\sim}11\%$. Correction for both effects (discrepancy in the $K-L$ color and imposition of the $L$-band cuts) is needed to bring the HLL01 disk fraction in agreement with ours.

\item {\bf \citet{Mamajek2009}.} He performs a literature compilation of heterogeneous sets of ages and disk fractions for 22 nearby young stellar clusters. His disk indicator is based on the combination of spectroscopic accretion and photometric IR excess signatures. Mamajek introduces the exponential decay formalism. For the compiled dataset he obtains a mean-lifetime of $\tau = 2.5$~Myr ($t_{1/2} = 1.7$~Myr). MYStIX/SFiNCs Siess00-based disk dissipation timescale of $t_{1/2} \sim 2$~Myr is consistent with that of Mamajek despite the heterogeneity of Mamajeks' dataset.

\item {\bf \citet{Fedele2010}.} They find that disk half-lives estimated based
on spectroscopic indicators of accretion ($\tau = 2.3$~Myr; $t_{1/2}=1.6$~Myr) are shorter than those based on
infrared excess ($\tau = 3.0$~Myr; $t_{1/2}=2.1$~Myr) to identify disks. The MYStIX/SFiNCs Siess00 result of $t_{1/2} \sim 2$~Myr agrees with their estimate of disk lifetime based on infrared excess.

\item {\bf \citet{Pecaut2012,Pecaut2016}.} They find that for the Scorpius--Centaurus OB association, the age estimates of low-mass (K- and M-type) PMS stars are systematically lower (by a factor of two) compared to those of higher-mass (G- and F-type) PMS and massive B-type main-sequence stars.  This age discrepancy is likely related to the problem of ``radius inflation'' in low-mass PMS stars (\S \ref{s:introduction}) and can be mitigated by introducing magnetic effects in PMS models \citep{Somers2015, Feiden2016}. For the Upper~Sco (US), Upper Centaurus-Lupus (UCL), and Lower Centaurus-Crux (LCC) sub-regions of the OB association, \citet{Pecaut2016} adopt HRD-based ages inferred for intermediate-mass G- and F-type PMS stars using the median age values among the output of four different, relatively modern PMS evolutionary models, including \citet{Dotter2008} and Baraffe15. Using these three clusters (US, UCL, and LCC), \citet{Pecaut2016} estimate disk dispersal half-life of $t_{1/2} = 3.3$~Myr ($\tau = 4.7$~Myr). 

The agreement with the MYStIX/SFiNCs half-life of $t_{1/2} \sim 3.5$~Myr (based on the Feiden16M time scale) is somewhat remarkable, considering the paucity of the Pecaut et al. cluster sample (only three clusters), as well as the following difference in disk indicators. As a disk indicator \citet{Pecaut2016} use {\it WISE} photometry, which goes significantly farther into the infrared (3.4--22~$\micron$), compared to {\it Spitzer}-IRAC employed by MYStIX/SFiNCs (3.6--8~$\micron$). By probing cooler, farther-out regions of disks, we would expect the Pecaut et al. study to arrive at longer disk dispersal timescales, but this has not occurred.

\item {\bf \citet{Ribas2014}.} The last possibility discussed above is supported by disparate estimates of disk lifetimes for
shorter wavelength (3.4--12~$\micron$; $\tau \la 3$~Myr) and longer wavelength (22--24~$\micron$; $\tau \ga 4$~Myr)
observations \citep{Ribas2014}, reflecting differential evolution for different
grain sizes and different regions of disks. \citet{Lada2006} classify several
objects in IC~348 as disk-bearing that are designated as disk-free in the
SFiNCs catalog. \citet{Lada2006} use 24~$\micron$ {\it Spitzer} data to help
identify disks, suggesting that the use of ($<$8~$\micron$) infrared data
as in the current work will fail to identify transition disks due to their
lack of a hot inner component.

\item {\bf \citet{Bell2013}.} They introduce new semi-empirical model isochrones to correct the aforementioned problem of radius inflation and systematically lower ages for low-mass stars derived using standard PMS models \citep{Pecaut2012,Pecaut2016}. Since a treatment of low-mass stellar ages is applied in Bell et al., we would expect to have our MYStIX/SFiNCs disk half-life of $t_{1/2} \sim 3.5$~Myr (based on the Feiden16M timescale), as well as that of \citet{Pecaut2016}, to be comparable with that of Bell et al. However, some complications arise. 

For 13 clusters, Bell et al. combine their new age estimates with {\it Spitzer}-based disk fractions compiled from the literature to obtain a disk dissipation timescale. Judging from their Figure~18, Bell et al. report a disk half-life of $\sim 5-6$~Myr. Two issues arise here. First, Bell et al. do not provide any formal exponential fits to the data. By applying the non-linear Gauss--Newton least-squares method to fit an exponential function $f_{disk} = f_0 \times e^{t/{\tau}}$ (with $f_0 = 100$\%) to the data given in their Figure 18, we derive a disk half-life of $t_{1/2} \sim 3$~Myr ($\tau \sim 4.3$~Myr). This is rather inconsistent with their own reported value of $\sim 5-6$~Myr. 

Second, several
regions are found in both the \citet{Bell2013} dataset and the combined
MYStIX and SFiNCs dataset.
These regions are Eagle Nebula, Lagoon Nebula, NGC~2264 (part of Rosette
Nebula), IC~5146, Cep~OB3b, and IC~348. Surprisingly, for the first four regions, these are $Age_{JX}$-Siess00 estimates and not $Age$-Feiden16M
that appear to agree fairly well with those reported by \citet{Bell2013}. MYStIX/SFiNCs $Age$-Feiden16M estimates are systematically higher than those of Bell et al.

For Cep~OB3b and IC~348, our $Age_{JX}$-Siess00 and $Age$-Feiden16M ages are in the 2.5--4.5~Myr range, as opposed to the Bell et
al. estimates of 6~Myr. While any number of factors affect age estimates, one partial explanation may be
the disparate distance estimates between \citet{Bell2013} and the current
work for Cep~OB3b and IC~348, while the distance estimates agree more closely
for the other four regions. The ${\sim}$20\% disagreement in distances for these two clusters is likely a major factor in explaining the age disparity. Tycho-{\it
Gaia} parallax distances for several objects in IC~348 confirm the SFiNCs
adopted distance of 300~pc, as opposed to the distance of 250~pc adopted by
\citet{Bell2013}. As for Cep~OB3b, Very Long Baseline Array parallax distances
for two objects in Cep~A---HW~2 and HW~9---are ${\sim}$700~pc
\citep{Moscadelli2009, Dzib2011}. Given the small (${\sim}$10~pc) projected
distance between Cep~A and Cep~OB3b \citep{Sargent1977}, we conclude that the
\citet{Bell2013} distance of 570~pc based on color--magnitude diagrams is likely
to be a substantial underestimation of the true distance. If the ages of
\citet{Bell2013} for these two clusters are replaced with their $Age_{JX}$-Siess00 values
from SFiNCs, the exponential half-life for the data shown in \citet{Bell2013}
Figure~18 further decreases from $t_{1/2} \sim 3$~Myr to $t_{1/2} \sim 2$~Myr (assuming 100\% initial
disk fraction). {\it Gaia} DR2 data will help to clarify these distance issues.
\end{enumerate}

\subsection{Suggestions for future work}

To conclude, we distill the discussions of Sections
\ref{s:results} and \ref{s:conclusions} into suggestions for future studies of disk
longevity.

\begin{enumerate}

\item Large datasets using homogeneously-derived cluster ages, based on PMS evolutionary models whose predictions are consistent with observations, are imperative for deriving reliable estimates of disk lifetimes and initial disk fraction. Our current work employing the largest cluster dataset used in a study of this kind to date with homogeneous sets of cluster ages based on the old (Siess00) and modern (MIST, Feiden16M) sets of evolutionary models shows a strong effect of the choice of PMS models on the disk longevity estimates.

As discussed in Section~\ref{s:noassume}, stronger
constraints on disk lifetimes and initial disk fractions will require a sample
that spans a range of stellar age longer than the characteristic timescale of
disk evolution. A larger age range will also allow for non-exponential
parametrizations of the data to be explored. A significant sample of stars
across all masses will be needed in order to resolve the question raised by our
results in Section~\ref{s:masses} of whether disks around high-mass stars evolve
similarly to those orbiting lower-mass stars.

\item Ensuring similar mass sensitivities for disk-bearing and disk-free sources
is important for deriving reliable estimates of disk longevity and initial disk
fraction, especially at older ages where disk fractions are low and subsequently
sensitive to small differences in numbers of sources. Although we are concerned about this issue, it did not prove to be so important in other studies such as \citet{Mamajek2009,Fedele2010,Ribas2014,Pecaut2016} except for HLL01. Since these previous studies focus mainly on nearby star-forming regions, it is possible that within the datasets employed, the disk-free and disk-bearing stellar samples have similar mass distributions.

The detection of disk-free
YSOs will be greatly facilitated by the use of X-ray data, as well as the use of
higher-sensitivity infrared instruments such as those offered by the James Webb
Space Telescope, especially for older objects with diminished X-ray
luminosities. The imposition of consistent stellar mass sensitivity limits for
disk-bearing and disk-free YSOs will be important for deriving strong
constraints on disk lifetimes and initial disk frequencies, and also for
exploring the role of stellar mass, cluster environment, and so on. Emission
from polycyclic aromatic hydrocarbons in rich star-forming regions may pose a
problem for ensuring consistent sensitivity toward disk-bearing and disk-free
stars even for high-mass stars. Ensuring consistent sensitivity at higher
masses may be important for calculating disk fractions among older systems
($>3-5$~Myr), where mass-dependent effects may emerge (based on the results of the
current work, they do not seem to emerge for systems younger than a few Myr; however, our disk fractions inferred for $\ga 2$~M$_{\odot}$ stars may be overestimated due to the diminished X-ray sensitivity in mid-B and mid-A type stars).

\item Future studies of disk fraction versus age should separately explore
multiple indicators of disks (while using otherwise homogeneous methods). In
particular, using a large range of infrared through sub-mm wavelengths, as well
as spectroscopic indicators, will help to determine how different regions of
disks evolve. Near-/mid-infrared (1--8~$\micron$) data is important for probing the
inner several AU of disks, providing constraints on the time available for the
in situ formation of Earth analogs (though the presence of near-/mid-infrared
excess does not necessarily indicate that a sufficient amount of dust for
building a planet is available). Longer wavelength data, on the other hand, can
provide insight into the evolution of the outer, cooler regions of disks, which
appear to evolve more slowly \citep{Ribas2014}.

\end{enumerate}

\section*{Acknowledgements}

We thank the referee for his/her very helpful comments. We thank K. Luhman, E. Mamajek, M. Pecaut, G. Somers, and R. Jeffries for stimulating discussions. The MYStIX project is now supported by the {\it Chandra} archive grant AR7-18002X. The SFiNCs project is supported at Penn State by NASA grant NNX15AF42G, {\it Chandra} GO grant SAO AR5-16001X, {\it Chandra} GO grant GO2-13012X, {\it Chandra} GO grant GO3-14004X, {\it Chandra} GO grant GO4-15013X, and the {\it Chandra} ACIS Team contract SV474018 (G. Garmire \& L. Townsley, Principal Investigators), issued by the {\it Chandra} X-ray Center, which is operated by the Smithsonian Astrophysical Observatory for and on behalf of NASA under contract NAS8-03060. The Guaranteed Time Observations (GTO) data used here were selected by the ACIS Instrument Principal Investigator, Gordon P. Garmire, of the Huntingdon Institute for X-ray Astronomy, LLC, which is under contract to the Smithsonian Astrophysical Observatory; Contract SV2-82024. This research has made use of NASA's Astrophysics Data System Bibliographic Services and SAOImage DS9 software developed by Smithsonian Astrophysical Observatory.

%%%%%%%%%%%%%%%%%%%%%%%%%%%%%%%%%%%%%%%%%%%%%%%%%%

%%%%%%%%%%%%%%%%%%%% REFERENCES %%%%%%%%%%%%%%%%%%

% The best way to enter references is to use BibTeX:

\bibliographystyle{mnras}
\bibliography{ms} % if your bibtex file is called example.bib

% Alternatively you could enter them by hand, like this:
% This method is tedious and prone to error if you have lots of references
%\bibliographystyle{mnras}
%\bibliography{example} % if your bibtex file is called example.bib
%%%\bibliography{Bibliography.bib}
%%%%%%%%%%%%%%%%%%%%%%%%%%%%%%%%%%%%%%%%%%%%%%%%%%

%\begin{figure*}
%\centering
%\includegraphics[angle=0.,width=170mm]{f1.pdf}
%\caption{Adaptively smoothed projected stellar surface densities of individual MYStIX clusters with a color-bar in units of observed stars per pc$^{2}$ (on a logarithmic %scale). The white circles mark the positions of stars with available $Age_{JX}$ estimates. The black polygons outline either the original full {\it Chandra}-ACIS-I fields of view for a single dominant cluster or the cutouts of the {\it Chandra} fields to separate the cluster of interest from other nearby clusters. \label{fig_maps_individual}}
%\end{figure*}

%%%%%%%%%%%%%%%%% APPENDICES %%%%%%%%%%%%%%%%%%%%%

%\appendix

%\section{Some extra material}

%If you want to present additional material which would interrupt the flow of the main paper,
%it can be placed in an Appendix which appears after the list of references.

%%%%%%%%%%%%%%%%%%%%%%%%%%%%%%%%%%%%%%%%%%%%%%%%%%

% Don't change these lines
\bsp	% typesetting comment
\label{lastpage}
\end{document}